\documentclass[useAMS,usenatbib,usegraphicx]{mn2e}
\usepackage{wasysym}
\title[Type III migration; Disc model]{Numerical
simulations of the type III migration:\\ I. Disc model and convergence
tests}
\author[A. Pepli\'nski et al.]{A. Pepli\'nski,$^1$\thanks{E-mail:
adam@astro.su.se} P. Artymowicz$^2$ and G. Mellema$^1$\\ $^1$Stockholm
University, AlbaNova University Centre, SE-106 91 Stockholm, Sweden\\
$^2$University of Toronto at Scarborough, 1265 Military Trail,
Toronto, Ontario M1C 1A4, Canada}
\begin{document}
\date{Accepted 0000 . Received 0000 ; in original form 0000 Month}
\pagerange{\pageref{firstpage}--\pageref{lastpage}} \pubyear{0000}

\maketitle

\label{firstpage} %
\begin{abstract}

We investigate the fast (type III)
migration regime of high-mass protoplanets orbiting in protoplanetary disks.
This type of migration is dominated by corotational torques. We study the
the details of flow structure in the planet's vicinity, the
dependence of migration rate on the adopted disc model, and the numerical
convergence of models (independence of certain numerical parameters
such as gravitational softening).

We use two-dimensional hydrodynamical simulations with adaptive mesh refinement,
based on the FLASH code with improved time-stepping scheme.  We perform global disk
simulations with sufficient resolution close to the planet, which is allowed to
freely move  throughout the grid. We employ a new type of equation of state in
which the gas temperature depends on both the distance to the star and planet,
and a simplified correction for self-gravity of the circumplanetary gas. 

We find that the migration rate in the type III migration regime depends
strongly on the gas dynamics inside the Hill sphere (Roche lobe of the planet)
which, in turn, is sensitive to the  aspect ratio of the circumplanetary disc.
Furthermore, corrections due to the gas self-gravity are necessary to reduce
numerical artifacts that act against rapid planet migration. Reliable numerical 
studies of Type III migration thus require consideration of both the thermal and
the self-gravity corrections, as well as a sufficient spatial resolution  and
the calculation of disk-planet attraction both inside and outside the  Hill
sphere. With this proviso, we find Type III migration to be a robust mode of
migration, astrophysically promising because of a speed much faster than
in the previously studied modes of migration.  
\end{abstract}

%

\section{Introduction}

In the standard model of planetary system formation giant planets form
through core accretion outside the ice condensation boundary
($4-5$~AU), where the availability of water ice allows planetary cores
to rapidly reach the critical mass of $\sim$10 Earth masses, beyond
which substantial gas accretion can occur
\citep{1996Icar..124...62P}. This theory provides a good fit to the structure of
the Solar System, but does not explain the presence of so-called `hot
Jupiters' (objects with minimum masses similar to or larger than 
Jupiter's mass, $M_{\jupiter}$, and semi-major
axes $a< 0.1$~AU) discovered in extrasolar planetary systems
(\citealt{1995Natur.378..355M}, \citealt{2000prpl.conf.1285M},
\citealt{2002ApJ...568..352V}). Since the in situ formation of these
objects is difficult both in the core accretion scenario and through
direct gravitational instability \citep{2001ApJ...563..367B}, the
inward migration of bodies and eccentricity pumping due to
planet-planet and planet-remnant disc interaction become important
parts of the new theory of planet formation.

The standard theory of planet migration
(\citealt{1979ApJ...233..857G,1980ApJ...241..425G};
\citealt{1993prpl.conf..749L}; \citealt{2000prpl.conf.1111L})
considers the Lindblad resonances and neglects the corotational
resonances, assuming a smooth initial density profile in the disc and a
small density gradient at the corotation radius. Tidal torques due to
the Lindblad resonances are believed to cause inward migration of a planet
(type I or type II, for low and high mass planets respectively) from
the region where it formed to the inner regions where many exoplanets are
observed. The type II migration times for giant planets are essentially viscous
time-scales of disks, thus relatively long \citep{1997Icar..126..261W}.

However, it was recently found that the corotational resonance can
modify the type I migration mode (\citealt{2006ApJ...652..730M}, 
\citealt{2006A&A...459L..17P}) or lead
to a new and very fast migration mode (type III, or run-away
migration) that depends strongly on the gas flow in the planet's
vicinity and does not have a predetermined direction
(\citealt{2003ApJ...588..494M}; \citealt{2004ASPC..324...39A};
\citealt{2005CeMDA..91...33P}; \citealt{2006AIPC..843....3A}). 
This type of migration was studied
numerically by \citet{2003ApJ...588..494M} who performed
two-dimensional simulations of a freely migrating planet and a steady
state migration with fixed migration rate $\dot a$ for a range
of the migration rates. Global, high resolution two- and
three-dimensional simulations of freely migrating planets were
performed by \citet{2005MNRAS.358..316D}. \citet{2005CeMDA..91...33P}
considered local shearing box simulations.

These numerical simulations showed that the planet's orbital evolution
depends strongly on the choice of the simulation parameters, e.g. grid
resolution \citep{2005MNRAS.358..316D}, softening of the planet
gravitation potential, etc. The reason for this dependence can be the
simplifications commonly used in the disc model. The most important
simplifications are the use of two-dimensional simulations, ignoring
self-gravity and the local-isothermal approximation, that imposes a
static temperature distribution in the planet's vicinity.

This is the first in a series of papers devoted to type III migration 
of high-mass protoplanets interacting with the protoplanetary disc. 
In this paper we study the dependence of the planet migration on the
applied disc model and we defer the description of the physics of type III migration
itself to \citet{PaperII} and \citet{PaperIII} (henceforth Paper~II and 
Paper~III). We concentrate on two
aspects: a modification of the local-isothermal approximation, that
allows an increase the temperature inside the Roche lobe, and a
correction of the gas acceleration (due to the gas self-gravity) that
forces the circumplanetary disc to move together with the planet.

In order to analyse this we study the gas flow in the planet's
vicinity by performing numerical hydrodynamical simulations in two
dimensions of a gaseous disc interacting with the star and one planet.
The planet is allowed to migrate due to disc-planet interaction and we
can study its orbital evolution in the non-steady state. Since type
III migration is so sensitive to the gas flow near the planet, we
include a careful analysis of the effects of various numerical
parameters which influence these flow patterns.

The layout of the paper is as follows. In Sections \ref{disc-model} and 
\ref{numerical_method} we give the basic equations, describe the disc 
model and the numerical method. Sections \ref{numerical_convergence_in} and 
\ref{sect_tor_mass} contain the convergence tests and 
a discussion of the effects of various modifications of disc model and 
numerical parameters for inward and outward migration. 
Finally in Section \ref{conclusions} we discuss 
their implications for numerical simulations of type III migration.


\section{Description of the physical model}

\label{disc-model}


\subsection{Disc model}

We adopt in our simulations a two-dimensional, infinitesimally thin
disc model and use vertically averaged quantities, such as the surface
mass density
\begin{equation}
\Sigma = \int \limits_{-\infty}^{\infty} \rho \, \rmn{d}z,
\end{equation}
where $\rho$ is the mass density. We work in the inertial reference
frame, in a Cartesian coordinate system $(x,y,z)$, also see
Sect.~\ref{sec_mesh_srt}. The plane of the disc and the star-planet
system coincides with the $z=0$ plane. The centre of mass of the star-
planet system is initially set at the origin of the coordinate
system. Since the star and the planet are allowed to migrate freely
due to the gravitational interaction with the disc, and the total angular
momentum of the whole system is not fully conserved due to the open
boundary conditions (see Section~\ref{boundary_cond}), the centre of
mass can move slowly away from the origin of the coordinate
system. The positions and masses of the star and the planet we denote
by $\bmath{r}_\rmn{S}$, $M_\rmn{S}$, $\bmath{r}_\rmn{P}$ and
$M_\rmn{P}$ respectively.

The gas in the disc is taken to be inviscid and
non-self-gravitating. The evolution of the disc is given by the
two-dimensional $(x,y)$ continuity equation for $\Sigma$ and the Euler
equations for the velocity components $\bmath{v} \equiv
(v_x,v_y)$. These equations can be written in conservative form as
\begin{eqnarray}
{\partial \Sigma \over \partial t} + \nabla(\Sigma \bmath{v}) &=& 0,\\
{\partial \Sigma \bmath{v} \over \partial t} + \nabla(\Sigma \bmath{v} \bmath{v}) 
+ \nabla P &=& -\Sigma \nabla \Phi,
\end{eqnarray}
where $P$ is two-dimensional (vertically integrated) pressure, and
$\Phi$ is the gravitational potential generated by protostar (subscript S) and
planet (subscript P)
\begin{equation}
\Phi = \Phi_\rmn{S} +\Phi_\rmn{P} = - {G M_\rmn{S} \over |\bmath{r} - 
\bmath{r}_\rmn{S}|} - {G M_\rmn{P} \over |\bmath{r} - \bmath{r}_\rmn{P}|}.
\end{equation}

We do not consider the energy equation, since we use a local
isothermal approximation (see Sect.~\ref{sec_eqn_of_state}).

The gravitational potential close to the
star and the planet is softened in the following way:
\begin{equation}
\widetilde{\Phi} = \left\{ \begin{array}{ccc}
{ -G  M \over r} &\hbox{; for  } \xi > 1 \\ 
\\
{ -G  M \over r_\rmn{soft}} (1.875 \xi^6 -  7 \xi^5  +7.875 \xi^4 + \\
 - 4.375\xi^2 +2.625)	&\hbox{; for  }   \xi \le 1
\end{array}\right.
\end{equation}
where $\xi = r/r_\rmn{soft}$ and $r_\rmn{soft}$ is the so-called smoothing 
length (or gravitational softening). 
The reason for using this formula instead of the standard one (where 
$r$ and  $r_\rmn{soft}$ are added geometrically) is to
remove the dependence of the Keplerian speed on the gravity smoothing
length for $r > r_\rmn{soft}$ in the disc surrounding the body. This
is especially important for simulations using a Cartesian grid
since the stellar gravity has to be softened too. Moreover it allows
for a bigger smoothing length around the planet without influencing
the outer part of the Roche lobe. This is necessary because the planet
is moving across the grid and its position with respect to cell
centres varies. As \citet{2003ApJ...589..556N} pointed out, in this
case, in order to avoid unphysical effects on the planet's trajectory
caused by close encounters with cell centres, the smoothing length
needs to be larger than half the dimension of the cell. In our
simulations $r_\rmn{soft}$ is at least a few times the cell size.

Unlike the standard formula for softening 
of the gravity, our formula corresponds to a spherical body with 
a finite radius given by $r_\rmn{soft}$. The internal structure of 
this body can be found using the Poisson equation and is given by the 
formula 
\begin{equation}
\label{eq3_dens_dist_inn}
\varrho_\rmn{soft} = \left\{ \begin{array}{ccc}
{ 0} &\hbox{; for  } \xi >1 \\
\\
{M \over {4 \pi r_\rmn{soft}^3}} (-78.75 \xi^4 + 210 \xi^3 + \\
-157.5 \xi^2 + 26.25) & \hbox{; for  } \xi \le 1
\end{array}\right.
\end{equation}
Since $r=r_\rmn{soft}$ is the ``surface'' of the body generating the 
gravitational potential, we will include the mass of the gas within 
the gravitational softening $M_\rmn{soft}$ to the effective planet 
mass. For more discussion see Sections~\ref{self-gravity} ,
\ref{conv_grav_mass_in} and~\ref{conv_grav_mass_out}.

Previous studies of migrating Jupiter-mass planets showed that an
important issue is the treatment of the torques arising from within
the Hill sphere (\citealt{2003ApJ...589..578N,
2005MNRAS.358..316D}). It is often assumed that these torques are
strong, but nearly cancel, and thus can be neglected. In addition it
has been argued that accelerating a planet through its own (bound)
envelope would be unphysical. In general, the exclusion of the torques
arising from within the Hill sphere is inconsistent, but can be
justified if the density distribution in the Hill sphere is (almost)
static and if the circumplanetary disc can be considered as a separate
system. As we show below, these assumptions are not necessarily
satisfied in the highly dynamic case of type III migration. 
The torques from the planet's vicinity need be taken into
account. Obviously, these torques depend strongly on the density
distribution near the planet, which in a numerical simulation will
depend on a combination of the resolution and the parameters
determining this density distribution. The gravitational softening
$r_\rmn{soft}$ is one of these parameters. In
Sect.~\ref{conv_grav_soft} we study the effects of gravitational
softening.

\begin{figure}
\includegraphics[width=84mm]{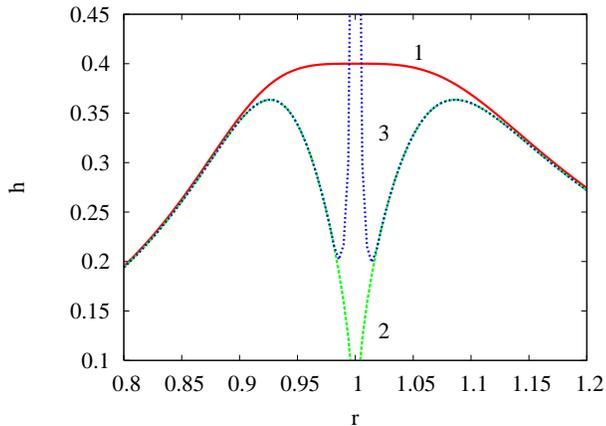}
\caption{The real disc aspect ratio calculated with respect to the
planet for two formulae for isothermal sound speed $c_\rmn{s}$.  Curve
1: EOS2 (Eq.~\ref{eq6_cs_star_planet}); curve 2: EOS1
(Eq.~\ref{eq6_cs_star}). Curve 3 shows the real aspect ratio for EOS2
after taking into account the gravitational softening (the result for
EOS1 is not affected by this).}
\label{f2_hpl}
\end{figure}

\subsubsection{Equation of state}

\label{sec_eqn_of_state}

Since a self-consistent calculation of the temperature is
prohibitively expensive, we adopt the usual local isothermal
approximation: the disc is treated as a system having a fixed
temperature distribution. In this case the equation of state has the
form
\begin{equation}
P=c_\rmn{s}^{2} \Sigma,
\end{equation}
$c_\rmn{s}$ being the local isothermal sound speed. The thermal
state of the fluid is given by $c_\rmn{s}$, which is usually assumed
to be a power law of the distance to the star $r_\rmn{s} = |\bmath{r}
- \bmath{r}_\rmn{S}|$. Assuming that the material inside a
circumstellar disc is in hydrostatic equilibrium and neglecting the
planet gravitational field, we can get the simple, widely used
formula  
\begin{equation}
c_\rmn{s}= H \Omega_\rmn{s},
\end{equation}
where $H$ is the disc scale height and $\Omega_\rmn{s}$ is the
Keplerian angular velocity in the circumstellar disc
\begin{equation}
\Omega_\rmn{s} = \sqrt{G M_\rmn{S} \over r_\rmn{s}^3}.
\end{equation}
This formula, for the constant disc opening angle, gives
\begin{equation}
\label{eq6_cs_star}
c_\rmn{s}= h_\rmn{s} r_\rmn{s} \Omega_\rmn{s},
\end{equation}
where $h_\rmn{s} = H/r_\rmn{s}$ is the disc aspect ratio with respect
to the star. In this case the Mach number of the flow in the disc is a
constant given by $1/h_\rmn{s}$. We will designate this choice as
EOS1 (equation of state 1).

For EOS1 the gas temperature is not modified by the presence of the
planet. Because of the gravitational field of the embedded planet, the
disc aspect ratio calculated with respect to the planet $h_\rmn{p} =
H/|\bmath{r} - \bmath{r}_\rmn{P}|$ can achieve very small values (see
Fig.~\ref{f2_hpl}), allowing the planet to collect large amounts of
material. Physically, we should expect a temperature increase in the
vicinity of a Jupiter-mass planet. Two-dimensional simulations
\citep{2003ApJ...599..548D} showed that an accreting Jupiter on a
constant orbit can have a circumplanetary disc with constant
$h_\rmn{p}$ ranging from 0.2 to 0.4, and three-dimensional simulations
show even higher values \citep{2006A&A...445..747K}. In the highly
dynamic case of a rapidly migrating planet $h_\rmn{p}$ can be expected
to attain similar values. For this reason we introduce a prescription
for the sound speed, which depends on the distance to both the star and
the planet
\begin{equation}
\label{eq6_cs_star_planet}
c_\rmn{s}={{h_\rmn{s} r_\rmn{s} h_\rmn{p} r_\rmn{p}} \over
{((h_\rmn{s} r_\rmn{s})^n+(h_\rmn{p} r_\rmn{p})^n)^{1/n}}}{\sqrt
{\Omega_\rmn{s}^2 +\Omega_\rmn{p}^2}},
\end{equation}
where $n$ is a non-dimensional parameter, 
$r_\rmn{p}= |\bmath{r} - \bmath{r}_\rmn{P}|$ is the distance to
the planet, and $\Omega_\rmn{p}$ is the Keplerian angular velocity in
the circumplanetary disc
\begin{equation}
\Omega_\rmn{p} = \sqrt{G M_\rmn{P} \over r_\rmn{p}^3}.
\end{equation}
This equation gives a constant disc aspect ratio in the circumstellar
disc $h_\rmn{s}$ far away from the planet, and a constant disc aspect
ratio in the circumplanetary disc $h_\rmn{p}$ in the planet's
vicinity. Parameter $n=3.5$ is chosen to smoothly join equations
(\ref{eq6_cs_star_planet}) and (\ref{eq6_cs_star}).  
We will refer to this approach as EOS2.

As pointed out above, the torques from within the Hill sphere are
important for type III migration, and will depend on the choice for
$h_\rmn{p}$ (as well as $r_\rmn{soft}$). Tests of how $h_\rmn{p}$
influences the convergence behaviour of our simulations are presented
in Section~\ref{conv_temp_prof}.


\subsubsection{Corrections for the gas self-gravity}
\label{self-gravity}

For low mass discs, it is usual to not include the effects of gas
self-gravity in numerical simulations. The argument is that
these effects are minor in not too massive disks, 
while the calculation of self-gravity
is particularly expensive. However, in the
case of type III migration the planet can collect a considerable
envelope with a mass equal to its own mass. The planet migrates
through interaction with the disc material, but without self-gravity,
its envelope will not do the same. This may lead to an artificial
increase of the planet's inertia, due to the discrepancy between the
planet position and the centre of mass of the planet's gaseous
envelope, and can result in an additional, non-physical force acting {\it against}\/ planet
migration \citep{2007prpl.conf..655P}.

\citet{2005MNRAS.358..316D} investigated type III migration and found
that increasing the spatial resolution of their simulations led to
gradually slower migration. Since higher resolution allows for more
mass to accumulate within the planet's Hill sphere, we believe that
their results reflect exactly the effect described above. 

Instead of calculating the full effect of self-gravity, we apply a
first order correction by forcing the planet and its gaseous envelope
to move together. This is done by modifying the acceleration of the
gas in a planet's vicinity ($\bmath{a}_\rmn{g}$), adding the planet
acceleration due to the torque from the gas $\bmath{a}_\rmn{T}$ to the
acceleration exerted on the gas by the star and the planet:
\begin{eqnarray} 
\bmath{a}_\rmn{g} = - {{G M_\rmn{S} (\bmath{r}-\bmath{r}_\rmn{S})} \over r_\rmn{s}^3} 
- {{G M^*_\rmn{P} (\bmath{r}-\bmath{r}_\rmn{P})} \over r_\rmn{p}^3} +\nonumber\\ 
+ \bmath{a}_\rmn{T} \max(0,1-{(r_\rmn{p}/r_\rmn{env})}^2),
\label{eq5_ag_env}
\end{eqnarray}
where $\bmath{r}_\rmn{S}$, $\bmath{r}_\rmn{P}$ are the positions of
the star and the planet, respectively. 
$M^*_\rmn{P}$ is taken to be either the planet mass $M_\rmn{P}$, or
$\widetilde M_\rmn{P}$, the planet mass plus all the gas mass within
the smoothing length $r_\rmn{soft}$ around the planet. The latter is
equivalent to assuming that the gas inside $r_\rmn{soft}$ attains the
spherical density distribution given by
Eq.~(\ref{eq3_dens_dist_inn}).

The last term gives the planet
acceleration multiplied by a function limiting the correction to the
planet's envelope using a parameter $r_\rmn{env}$. $r_\rmn{env}$
should be of order the size of the circumplanetary disc with
$R_\rmn{H} > r_\rmn{env} > r_\rmn{soft}$. We find that $r_\rmn{env} =
0.5 R_\rmn{H}$ removes the artificial increase of the planet's
inertia, provided the density distribution inside $r_\rmn{env}$ is
relatively symmetric and smooth, as is the case in
our simulations.

Notice that $r_\rmn{env}$ and $r_\rmn{soft}$ are independent parameters.
$r_\rmn{soft}$ gives the position 
of the ``surface'' of the protoplanet (radius where $\widetilde{\Phi}$ 
starts to differ from the  point-mass potential), whereas 
$r_\rmn{env}$ gives the size of the region that dynamically belongs to 
the planet and should follow the planet in its radial motion. 
Our correction 
for the gas acceleration allows to reduce the non-physical 
eccentricity of the orbits in the circumplanetary disc driven by the planet's 
radial motion.
The $r_\rmn{env}$ is thus a third parameter influencing
the density distribution near the planet (the others being the
gravitational smoothing, $r_\rmn{soft}$, and the disc aspect ratio in
the circumplanetary disk, $h_\rmn{p}$). The effects of $r_\rmn{env}$
on the numerical convergence for type III migration are
described in sections~\ref{conv_self_grav}
and~\ref{conv_grav_soft_out}. The differences between using
$M^*_\rmn{P} = M_\rmn{P}$ and $M^*_\rmn{P} =\widetilde M_\rmn{P}$ are
studied in Sect.~\ref{conv_grav_mass_in}.

We stress that the described method is a crude approximation
introduced to remove the non-physical effects from the planet's
orbital evolution. It does not allow any detailed 
study of the real flow of self-gravitating gas inside the Roche lobe or of 
the structure of the planet's gaseous envelope, 
which are beyond the scope of this paper.


\subsubsection{Accretion onto the planet}

\label{desc_mod_accr}

The orbital evolution can also be affected by gas accretion onto the
planet. This can be dealt with either by using $M^*_\rmn{P}=\widetilde M_\rmn{P}$
instead of $M_\rmn{P}$ in Eq.~\ref{eq5_ag_env}, or by removing matter
from the planet's neighbourhood $|\bmath{r}-\bmath{r}_\rmn{P}|<
r_\rmn{acc}$, and adding it to $M_\rmn{P}$. In the latter case, we
also add the momentum of the accreted gas to that of the planet's.
When removing gas from the disc, this is done after each integration
step according to the formula
\begin{equation}
\Delta \Sigma = \max \left(0,{dt \over \tau_\rmn{acc}}\max \left(0,1-{|\bmath{r}-
\bmath{r}_\rmn{P}|^2 \over r_\rmn{acc}^2}\right) (\Sigma - \overline\Sigma
)\right),
\end{equation}
where $\tau_\rmn{acc}$ is the accretion time-scale and
$\overline\Sigma$ is an average surface density in the region defined
by $r_\rmn{acc}<|\bmath{r}-\bmath{r}_\rmn{P}|< 2r_\rmn{acc}$. The size
of the accretion region $r_\rmn{acc}$ is defined by the size of the
smoothing length $r_\rmn{soft}$, i.e.\ the size of the
gravitational source. In simulations where we explicitly 
include accretion, we use
\begin{equation}
 r_\rmn{acc} = 0.5 r_\rmn{soft},
\end{equation}
which is an order of magnitude smaller than the Roche lobe size. The 
effects of the different ways to deal with accretion are studied in
Sect.~\ref{conv_grav_mass_in}.


\subsection{Equation of motion for the star and the planet}

The goal of the current study is to investigate the orbital evolution
of the planetary system due to the gravitational action of the disc
material. Since the calculations are done in the inertial reference
frame, the equation of motion for the star and the planet have the
simple form:
\begin{eqnarray}
\ddot {\bmath{r}}_\rmn{S} = - {G M_\rmn{P} (\bmath{r}_\rmn{S} -
\bmath{r}_\rmn{P}) \over {|\bmath{r}_\rmn{P} - \bmath{r}_\rmn{S}|}^3}
- \int\limits_{M_\rmn{D}} {G (\bmath{r}_\rmn{S} - \bmath{r})
\, dM_\rmn{D}(\bmath{r}) \over {|\bmath{r} - \bmath{r}_\rmn{S}|}^3}, \\
\ddot {\bmath{r}}_\rmn{P} = - {G M_\rmn{S} (\bmath{r}_\rmn{P} -
\bmath{r}_\rmn{S}) \over {|\bmath{r}_\rmn{P} - \bmath{r}_\rmn{S}|}^3}
- \int\limits_{M_\rmn{D}} {G (\bmath{r}_\rmn{P} - \bmath{r})
\, dM_\rmn{D}(\bmath{r}) \over {|\bmath{r} - \bmath{r}_\rmn{P}|}^3}.
\end{eqnarray}
In both cases the integration is carried out over the disc mass
$M_\rmn{D}$ included inside the radius $r_\rmn{disc}$ (thus removing
the corners of the Cartesian grid from consideration).

\begin{figure}
\includegraphics[width=84mm]{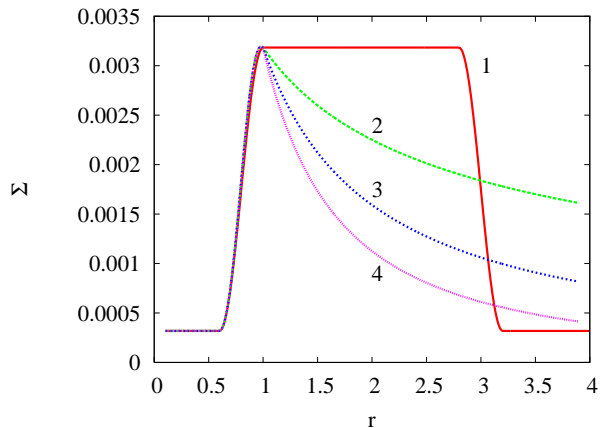}
\caption{The initial surface density profile for 
$\alpha_\rmn{\Sigma}$ ranging from $0.0$ to $-1.5$ (outward migration case).}
\label{fn_init_dens_prof}
\end{figure}


\subsection{Simulation setup}

\label{sec_setup-desc}

In the simulations we adopt non-dimensional units, where the sum of
star and planet mass $M_\rmn{S} + M_\rmn{P}$ represents the unit of
mass. The time unit and the length unit are chosen to make the
gravitational constant $G=1$. This makes the orbital period of
Keplerian rotation at a radius $a=1$ around a unit mass body equal to $2
\pi$. However, when it is necessary to convert quantities into
physical units, we use a Solar-mass protostar $M_\rmn{S}
=M_{\astrosun}$, a Jupiter-mass protoplanet $M_\rmn{P}=M_{\jupiter}$,
and a length unit of $5.2$~AU. This makes the time unit equal to
$11.8/2\pi$ years.

In all the simulations the grid extends from $-4.0$ to $4.0$ in both
directions around the star and planet mass centre. This corresponds to
a disc region with a physical radius of $20.8$ AU.


\subsubsection{Initial conditions}

\label{setc_init_cond}

The initial surface density $\Sigma$ profile is given by a modified power law:
\begin{equation}
\Sigma = \psi(r_\rmn{c}) \Sigma_\rmn{0}  (r_\rmn{c}/r_0)^{\alpha_\rmn{\Sigma}},
\end{equation}
where $r_\rmn{c} = |\bmath{r} - \bmath{r}_\rmn{C}|$ is the distance to
the mass centre of the planet-star system, $r_0$ is a unit distance, 
and $\psi$ is a function
that allows introducing a sharp edges in the disc (see
Fig.~\ref{fn_init_dens_prof}).

We characterize the disc mass by the disc to the primary mass ratio
\begin{equation}
\mu_\rmn{D} = {{\Sigma(r_0) \pi r_0^2} \over 
{M_\rmn{S}}} = {{\Sigma_\rmn{0} \pi r_0^2} \over {M_\rmn{S}}}.
\end{equation}
In the simulations $\mu_\rmn{D}$ ranges from $0.001$ to $0.01$. For
the Minimum Mass Solar Nebula (MMSN) $\mu_\rmn{D}=0.00144$
(for $\alpha_\rmn{\Sigma}=-3/2$). We investigate different density
profiles by changing $\alpha_\rmn{\Sigma}$ from $-1.5$ to $0.0$. 

Since we focus on type III migration only and do not analyse the
problem of orbital stability inside a gap, we do not introduce the
planet smoothly nor keep it on a constant orbit for the time needed to
create a gap. For most cases with $\alpha_\rmn{\Sigma} < 0.0$ the
density gradient given by the initial profile is sufficient to start
rapid inward migration. In the case of outward migration we start
migration by introducing an additional density jump at the planet's
position.

The planet mass is taken to be $M_\rmn{P}/M_\rmn{S} = 0.001$ (i.e. one
Jupiter mass, $M_{\jupiter}$, for a one-solar-mass star). The planet
starts on a circular orbit of semi-major axis equal $3.0$ and $0.8$
for the inward and outward migration case respectively.

The aspect ratio for the disc with respect to the star is fixed at
$h_\rmn{s} = 0.05$, whereas the circumplanetary disc aspect ratio
$h_\rmn{p}$ is taken from the range $0.2$ to $0.6$.

The smoothing length of the stellar potential, $r_\rmn{soft}$, is
taken to be $0.5$. Unless otherwise noted, for the planet the standard value is $0.33 R_\rmn{H}$, where
$R_\rmn{H} = a{[M_\rmn{P}/(3M_\rmn{S})]}^{(1/3)}$, the Hill
radius.
The corresponding size of the envelope $r_\rmn{env}$ in
Eq.~(\ref{eq5_ag_env}) is set to $r_\rmn{soft}$ or $0.5
R_\rmn{H}$.

If accretion onto the planet is included, we use an accretion time-scale 
$\tau_\rmn{acc} = 10 \pi$ and $ r_\rmn{acc} = 0.5 r_\rmn{soft}$.


\subsubsection{Boundary conditions}

\label{boundary_cond}
To stop reflections off the boundary, we use an outflow-inflow
boundary condition with a so-called killing wave zone.  This zone
extends from radius $3.65$ to $3.9$. In this region the solution of
the Euler equations for $X$ ($X$ stands for the surface density or the
velocity) is changed after each time-step by
\begin{equation}
\Delta X = {{dt} \over {\tau_\rmn{d}}} (X - X_\rmn{0}) \phi(r),
\end{equation}
where $X_\rmn{0}$ is the initial condition (disc in sub-Keplerian
rotation), $\tau_\rmn{d}$ is the damping time-scale, and $\phi(r)$
is a function increasing from $0$ at the inner boundary of the
killing wave zone to $1$ at the outer boundary of the killing wave
zone. Outside the radius $3.9$ the solution $X$ is replaced by
$X_\rmn{0}$. In our simulations we use $\tau_\rmn{d}= 18$.

On our Cartesian grid the disc cannot be treated as an isolated system
and some mass and angular momentum flow through the boundary is
unavoidable. However, the losses are usually relatively small.

\begin{figure*}
\includegraphics[width=84mm]{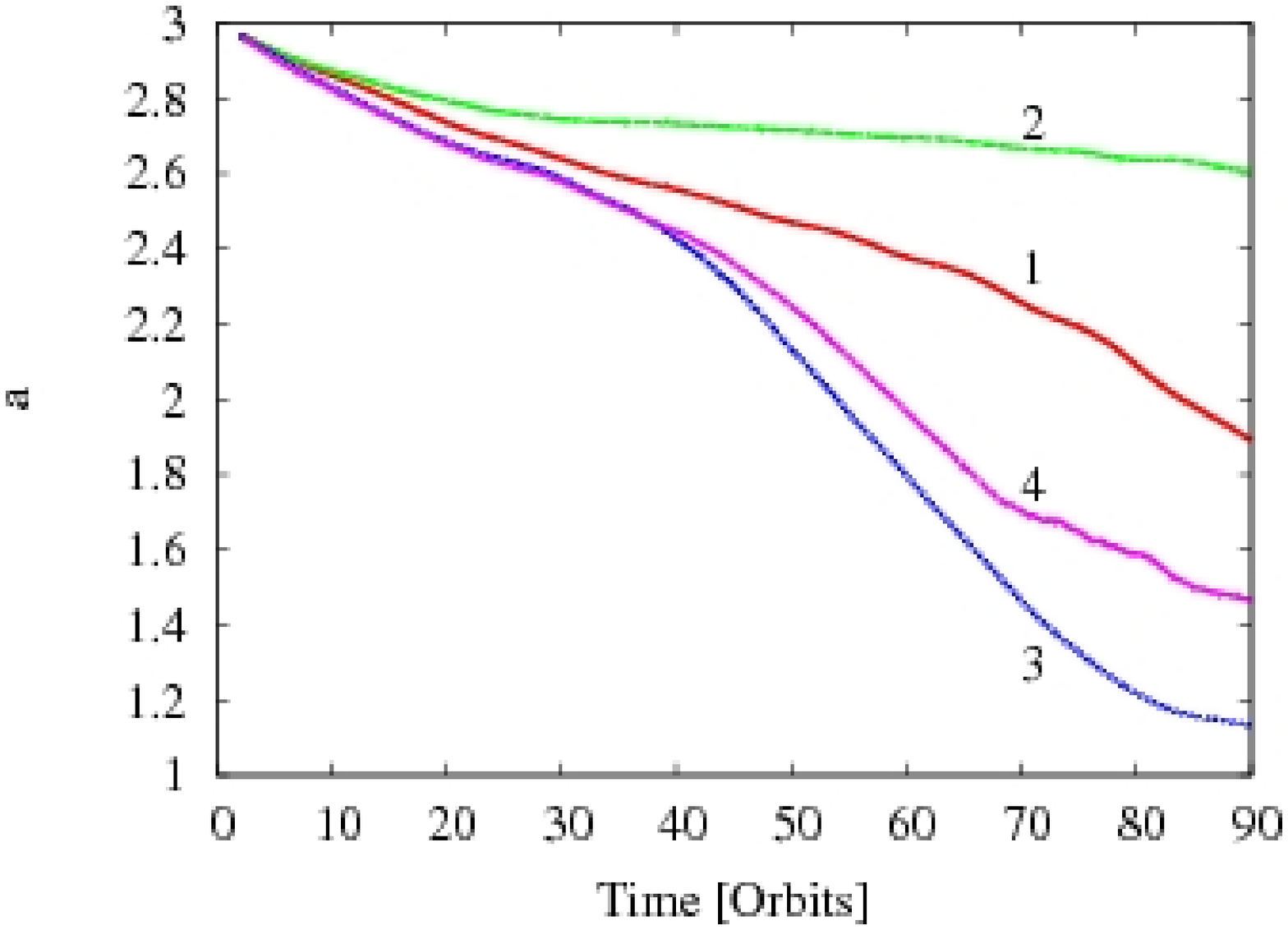}
\includegraphics[width=84mm]{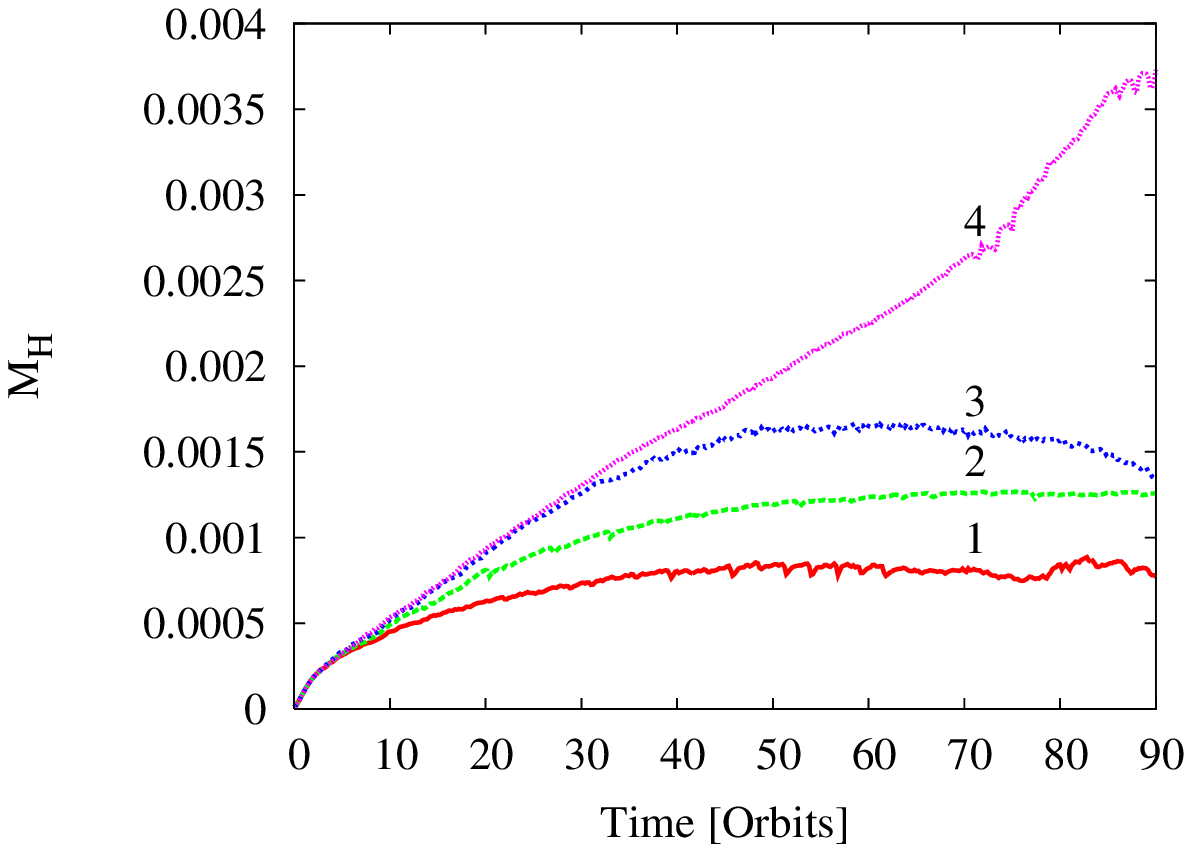}
\caption{Convergence test for the inward migrating Jupiter. The left
panel shows the time evolution of the planet's semi-major axis $a$,
the right panel the mass of the gas inside a Hill sphere $M_H$.
Curves 1 and 2 correspond to simulations with 4 and 5 levels of
refinement and $r_\rmn{env} = 0$ (planet acceleration is not added to
the gas acceleration); curves 3 and 4 correspond to simulations with 4
and 5 level of refinement and $r_\rmn{env} = 0.5 R_\rmn{H}$. The time
unit is the orbital period of the body circulating at $a = 1$.
The planet mass $M_\rmn{P} = 0.001$.}
\label{f2_ac2}
\end{figure*}


\section{Description of the numerical method}

\label{numerical_method}


\subsection{Code}

We adopted the {\it FLASH} hydro code version 2.3 written by the {\it
FLASH Code Group} from the Centre for Astrophysical Thermonuclear
Flashes at the University of Chicago
\footnote{http://flash.uchicago.edu} in 1997.

{\it FLASH} is a modular, adaptive-mesh, parallel simulation code
capable of handling general compressible flow problems. It is designed
to allow users to configure initial and boundary conditions, change
algorithms, and add new physics modules. It uses the {\it PARAMESH}
library \citep{2000CPC.126..330} to manage a block-structured adaptive
mesh, placing resolution elements only where they are needed
most. {\it PARAMESH} consists of a set of subroutines which handle
refinement/derefinement, distribution of work to processors, guard
cell filling, and flux conservation. It also uses the Message-Passing
Interface (MPI) library to achieve portability and scalability on
parallel computers. For our purpose, the code is used in the pure
hydrodynamical mode in two dimensions, and the adaptive mesh is used
to achieve high resolution around the planet (see Sect.~\ref{sec_mesh_srt}).

Euler's equations are solved using a directionally split version of
the {\it Piecewise-Parabolic Method} ({\it PPM},
\citealt{1984JCP.54..174}). It represents the flow variables inside
the cell with piecewise parabolic functions.  This method is
considerably more accurate and efficient than most formally
second-order algorithms.

To solve the equations of motion for the star and the planet we use a
4-th order Runge-Kutta method. To keep the second order accuracy in
time {\it PPM} needs information about the gravitational field from
the beginning and the end of the time-step. Therefore we perform the
integration of the planet's orbit in two steps. First we integrate
using the force exerted by gas taken from the previous hydro-step,
This provides {\it PPM} with the approximate position of the planet at
the end of the current time-step. Second, after updating the
hydrodynamic quantities, we correct the planet position using a linear
interpolation between the old and new values of acceleration.


\subsection{Mesh and grid structure}

\label{sec_mesh_srt}

High resolution is needed to calculate accurately the gas flow in the
planet's vicinity. For this purpose we use the Adaptive Mesh
Refinement ({\it AMR}) module included in {\it FLASH}. {\it AMR}
allows us to change the grid structure during the simulation, and
gives the possibility for the refined region to follow the planet's
motion. The resolution is increased by a factor of 2 between
refinement levels. Our simulations use a lowest resolution mesh of
$800$ cells in each direction, and a square region around the
planet is refined. The maximal cell size in the disc (lowest level of
refinement) is about 1\% of the smallest value of the planet's
semi-major axis. The cell size close to planet is at least 1.8\% of
Hill sphere radius (corresponding to 4 levels of refinement).

We use a Cartesian grid for our calculations. This choice was made
because the more usual cylindrical (co-rotating) grid geometry has few
benefits for the case of a rapidly migrating planet, and also suffers
from variable (physical) resolution with radius. However, this does
not mean that a Cartesian grid is problem free. The most important 
problem is the diffusivity of the code visible in \citet{2006MNRAS.370..529D}, 
however the time-scale of this process is relatively long and does 
not influence the rapid migration. It is discussed 
in Appendix~\ref{app_num_diff}.


\subsection{Multi-level time integration}

In order to reduce the computational time we extended {\it FLASH} with
the option of different time-steps for different refinement levels.
The algorithm used is very similar to nested-grid technique presented
in \citet{2002A&A...385..647D}. Coarser blocks can use longer time-steps 
than finer blocks, but all blocks at a given resolution have
to use the same time-step. In this case a time-step is a monotonic
function of the grid resolution (in {\it FLASH} all time-steps have to
be power of 2 multiples of the time-step being used on the finest
block). This resolution dependent time-step can save a lot of computer
time, but perhaps more importantly reduces the numerical diffusion in
blocks at coarser refinement levels, which need a fewer number of
time-steps to calculate a given evolution time.


\section{Numerical convergence}

\label{numerical_convergence_in}

The problem of type III migration is numerically challenging since we
have to deal with the complex evolution of the co-orbital flow in a
partially gap opening regime, close to the planet. Clearly one can
expect the results to depend on the disk model close to the planet,
and in this section we investigate how the convergence behaviour
(dependence on resolution) depends on our choices for the correction
for self-gravity ($r_\rmn{env}$), circumplanetary disc aspect ratio
($h_\rmn{p}$), and gravitational softening ($r_\rmn{soft}$).
Here we discuss the inward migration case only, 
but the conclusion are valid for the outward directed migration too.

\citet{2005MNRAS.358..316D} showed that it is hard to achieve
numerical convergence for these types of flows. They found the
migration to be highly dependent on the torque calculation
prescription, and on the mesh resolution (see their Fig.~8 and~9).  We
can qualitatively reproduce their results when using EOS1, and for
$r_\rmn{env} = 0$ and $M^*_\rmn{P} = M_\rmn{P}$
(Eq.~\ref{eq5_ag_env}), see curves 1 and 2 in Fig.~\ref{f2_ac2}.
In this case higher resolution allows more mass to
accumulate in the planet's vicinity, and the lack of any self-gravity
allows this mass to increase the planet's inertia, slowing down its
migration.

\begin{figure}
\includegraphics[width=84mm]{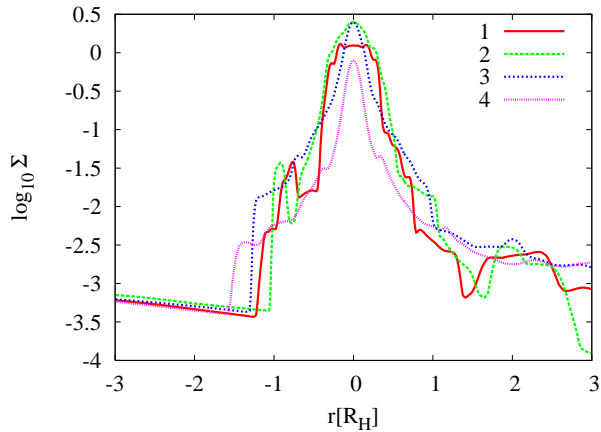}
\caption{Surface-density radial cut through the Roche lobe at the
planet position for different temperature profiles. Curve 1
corresponds to EOS1 (Eq.~(\ref{eq6_cs_star}); curves 2, 3 and 4
to EOS2 (Eq.~(\ref{eq6_cs_star_planet}) with $h_\rmn{p}$ equal
$0.2$, $0.3$ and $0.4$ respectively.}
\label{f4_srl_cut}
\end{figure}


\subsection{Correction for gas self-gravity}

\label{conv_self_grav}

\begin{figure}
\includegraphics[width=84mm]{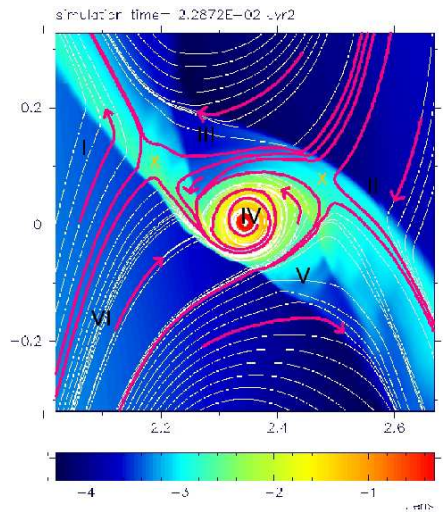}
\caption{The flow in the planet's proximity for the rapid inward
 migration. In this case the flow symmetry is broken and the big ammount
 of the mass is accumulated in the planet's proximity. The plot shows
 the surface density distribution and the flow lines in the frame
 co-moving with the planet and covers a square region of the size of
 $4R_\rmn{H}$. The colour scale is logarithmic. The pink lines show the
approximate borders of the different regions in the disc: inner disc
I, outer disc II, horseshoe region III, circumplanetary disc IV,
co-orbital flow transferring the gas from the inner to the outer disc
V and the gas stream entering the circumplanetary disc VI. The arrows
show the direction of the flow.}
\label{f4_flow_rl}
\end{figure}

As explained in Sect.~\ref{self-gravity} neglecting self-gravity is
likely to be one cause of this bad convergence behaviour. To explore
this further we use the correction for self-gravity described in
Sect.~\ref{self-gravity}. The key parameter of this correction is
$r_\rmn{env}$, the size of the region where the gas is dynamically
connected to the planet. To establish its value, we performed a series
of simulations with different resolutions (3, 4, and 5 refinement
levels) and different values of $r_\rmn{env}$: $0.0$ (no correction),
$r_\rmn{soft}$ (the smoothing length for gravity) and $0.5
R_\rmn{H}$. The last value is bigger than the estimated radius of the
circumplanetary disc. From the simulations we know that the
circumplanetary disc extends up $(0.3-0.4) R_\rmn{H}$ (see
Fig.~\ref{f4_flow_rl}). We can expect $r_\rmn{env}$ to be bigger than
the radius of the circumplanetary disc, since Eq.~(\ref{eq5_ag_env}) does not define a sharp edge to the planet's
envelope.  Some of the results of these tests are presented in
Fig.~\ref{f2_ac2}. In these simulations we used a constant
$M^*_\rmn{P} = M_\rmn{P}$.

We see that the correction reduces the `additional inertia' and
increases the migration rate, even though the mass of the gas
accumulated in the Hill sphere increases significantly (right
panel). For $r_\rmn{env} = r_\rmn{soft}$ the correction region is
smaller than the real size of the planet envelope, and we find results
similar to curves 1 and 2. The convergence is much better for
$r_\rmn{env} = 0.5 R_\rmn{H}$ (curves 3 and 4), but even there we have
not reached complete convergence with 5 refinement levels. This is
caused by the mass accumulation in the circumplanetary disc. Comparing
the simulations with 4 and 5 refinement levels (curves 3 and 4
respectively) we can see that the planet migrates faster after the
mass accumulation in the circumplanetary disc stops. As the amount of
mass accumulation depends on the assumed thermal structure of the
circumplanetary disc (here taken to follow EOS1), we will explore this
point more in the next section.

One may consider excluding the Roche lobe interior from the torque
calculation as another solution to the artificial inertia problem 
\citep{2003ApJ...588..494M}. However, the gas flow in the planet
vicinity is very variable and it is impossible to define a single
cutting radius to remove the region dynamically connected to the
planet from the torque calculation and at the same time keep all the
flow-lines that belong to the corotational flow. Moreover, in the case
of the fast migration, the interior of the Roche lobe is not separated
from the corotational flow and we cannot neglect the gas inflow into
the circumplanetary disc (Fig.~\ref{f4_flow_rl}).

\begin{figure*}
\includegraphics[width=84mm]{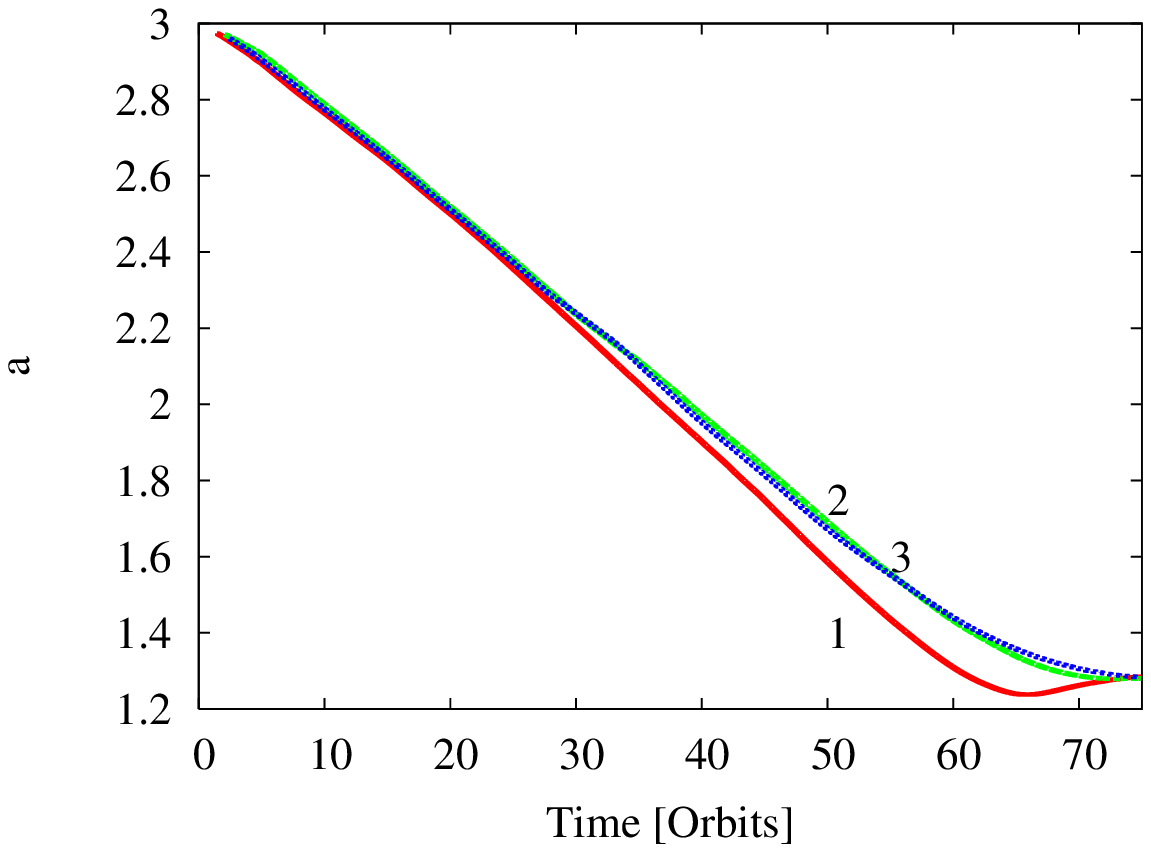}
\includegraphics[width=84mm]{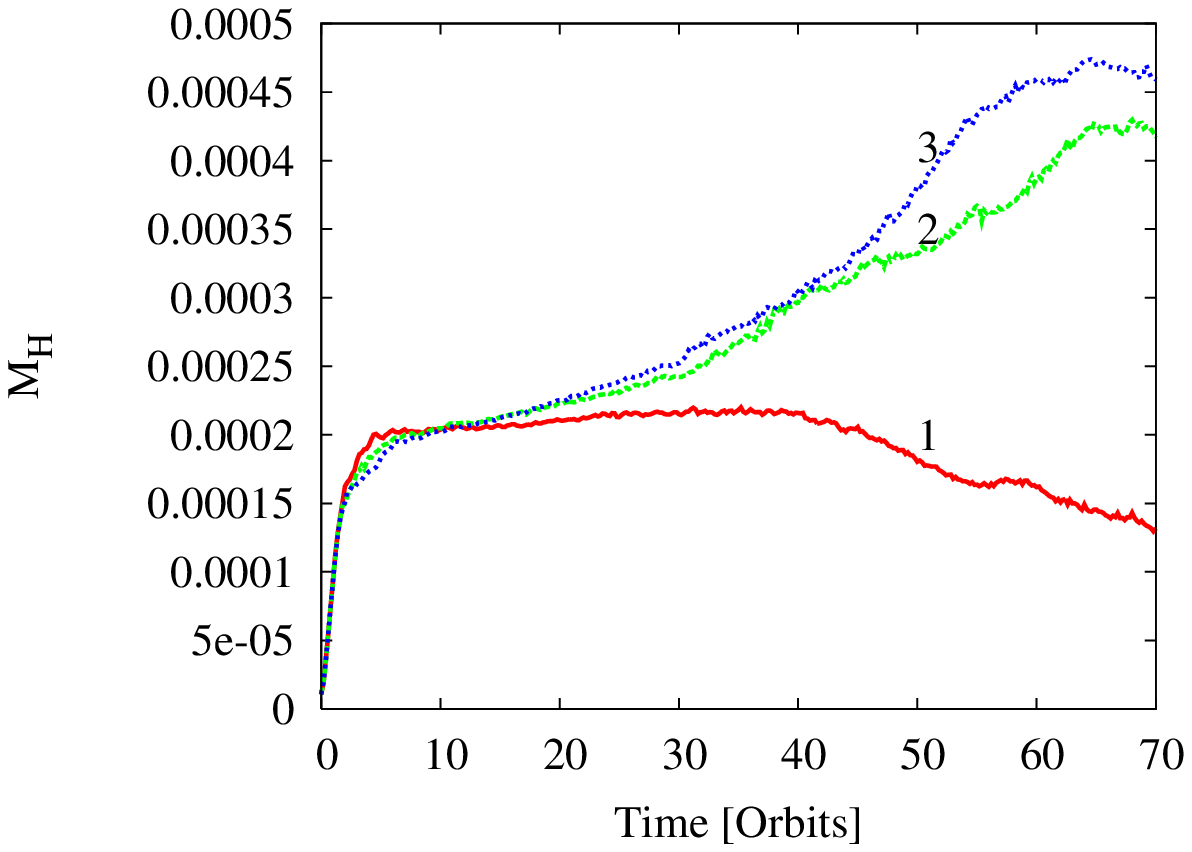}
\caption{Convergence test for the inward migrating Jupiter with the
disk aspect ratio $h_\rmn{p}=0.4$. 
The left panel shows the time evolution of the planet's semi-major axis $a$,
the right panel the mass of the gas inside a Hill sphere $M_H$.
Curves 1, 2 and 3 correspond to 3,
4 and 5 levels of refinement. The planet mass $M_\rmn{P} = 0.001$.}
\label{f7_af4}
\end{figure*}

\begin{figure*}
\includegraphics[width=84mm]{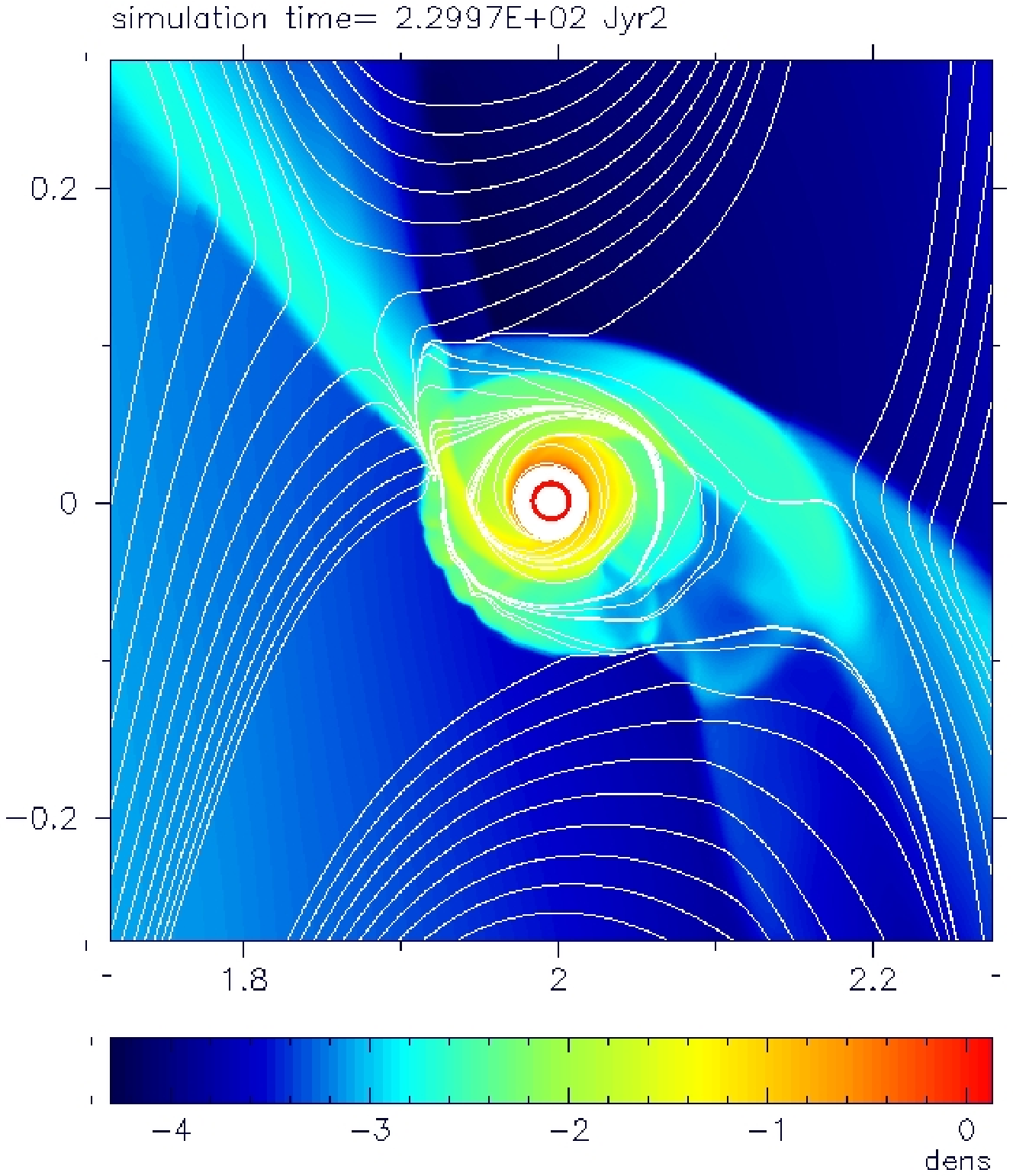}
\includegraphics[width=84mm]{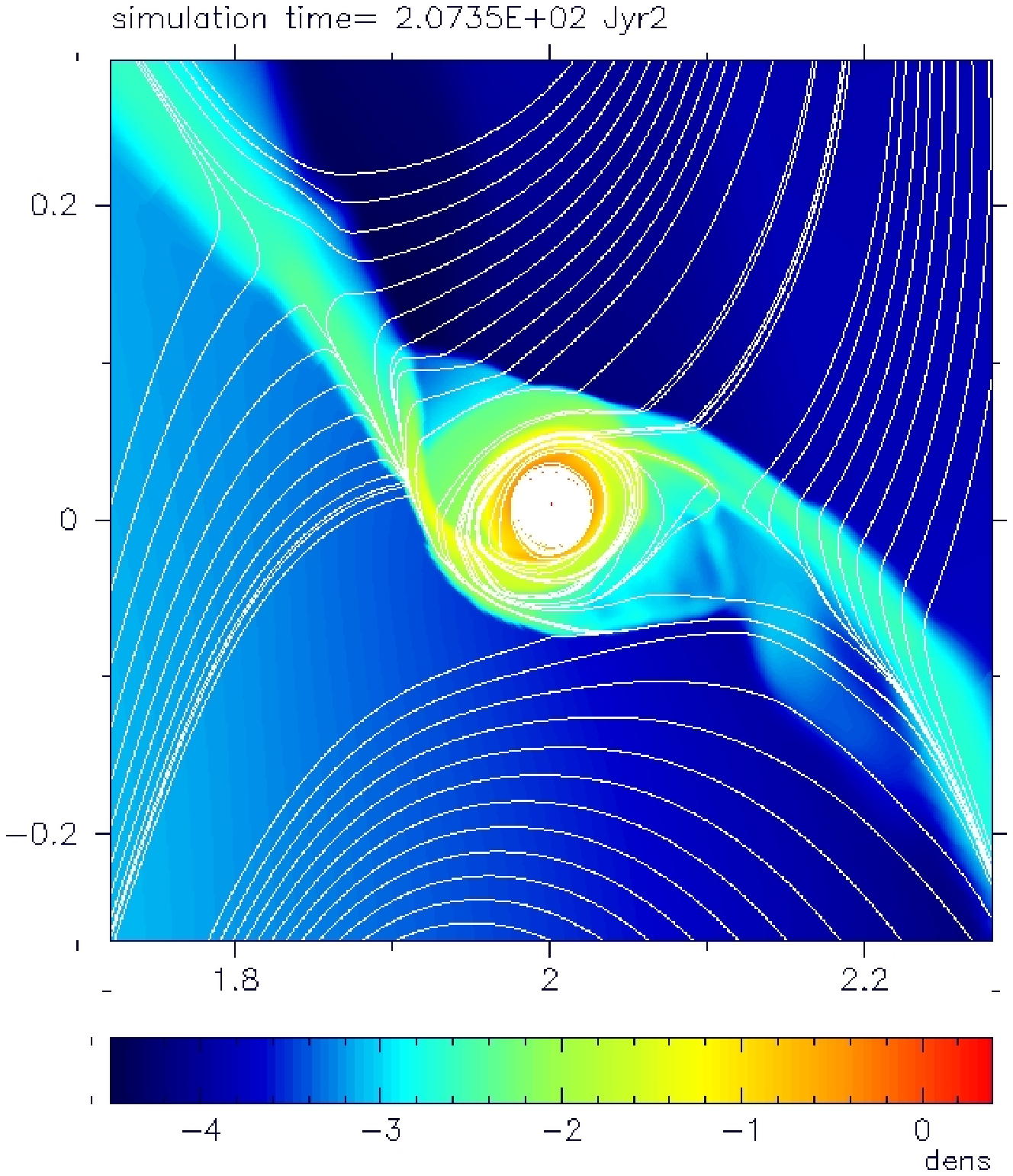}
\includegraphics[width=84mm]{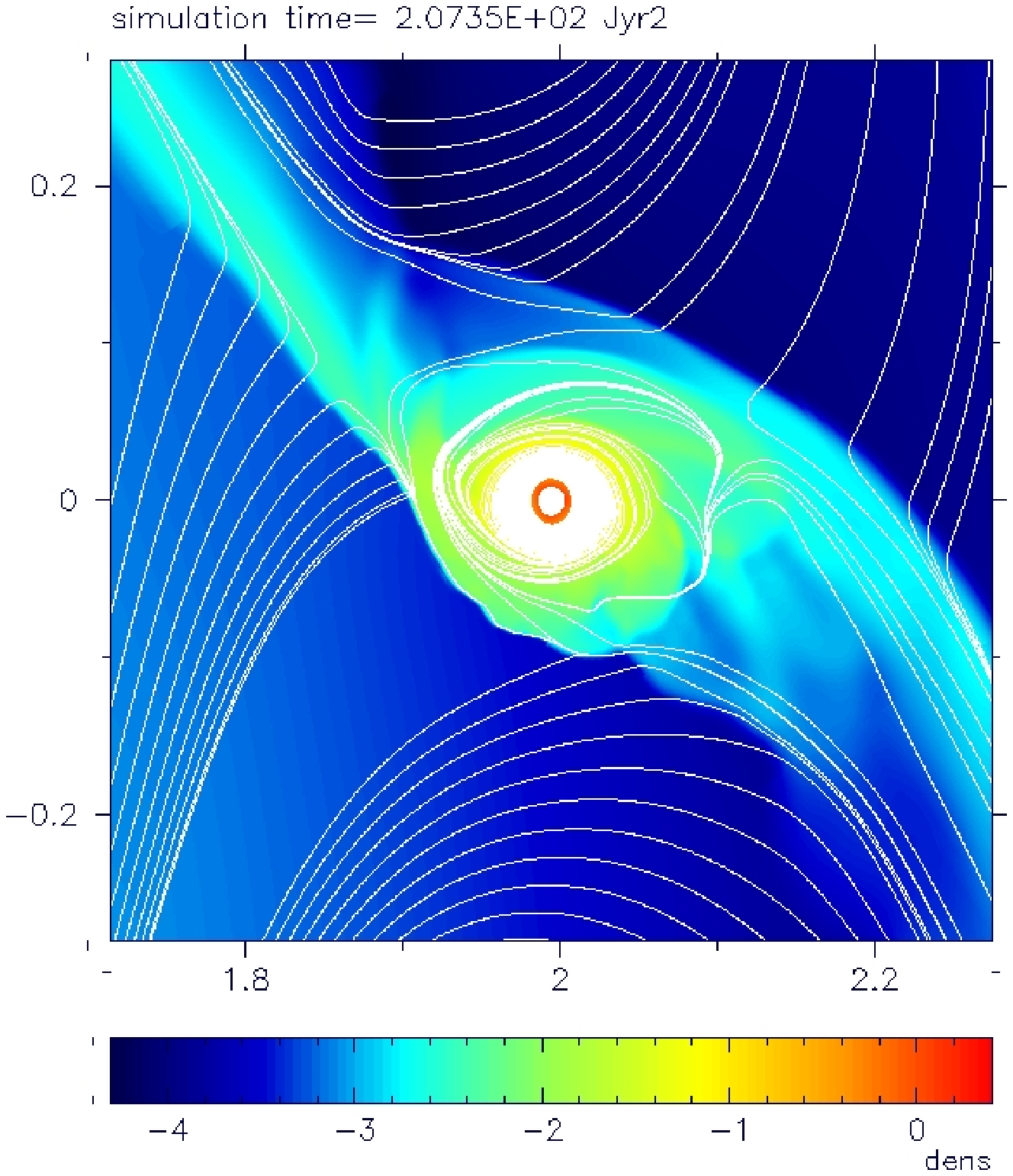}
\includegraphics[width=84mm]{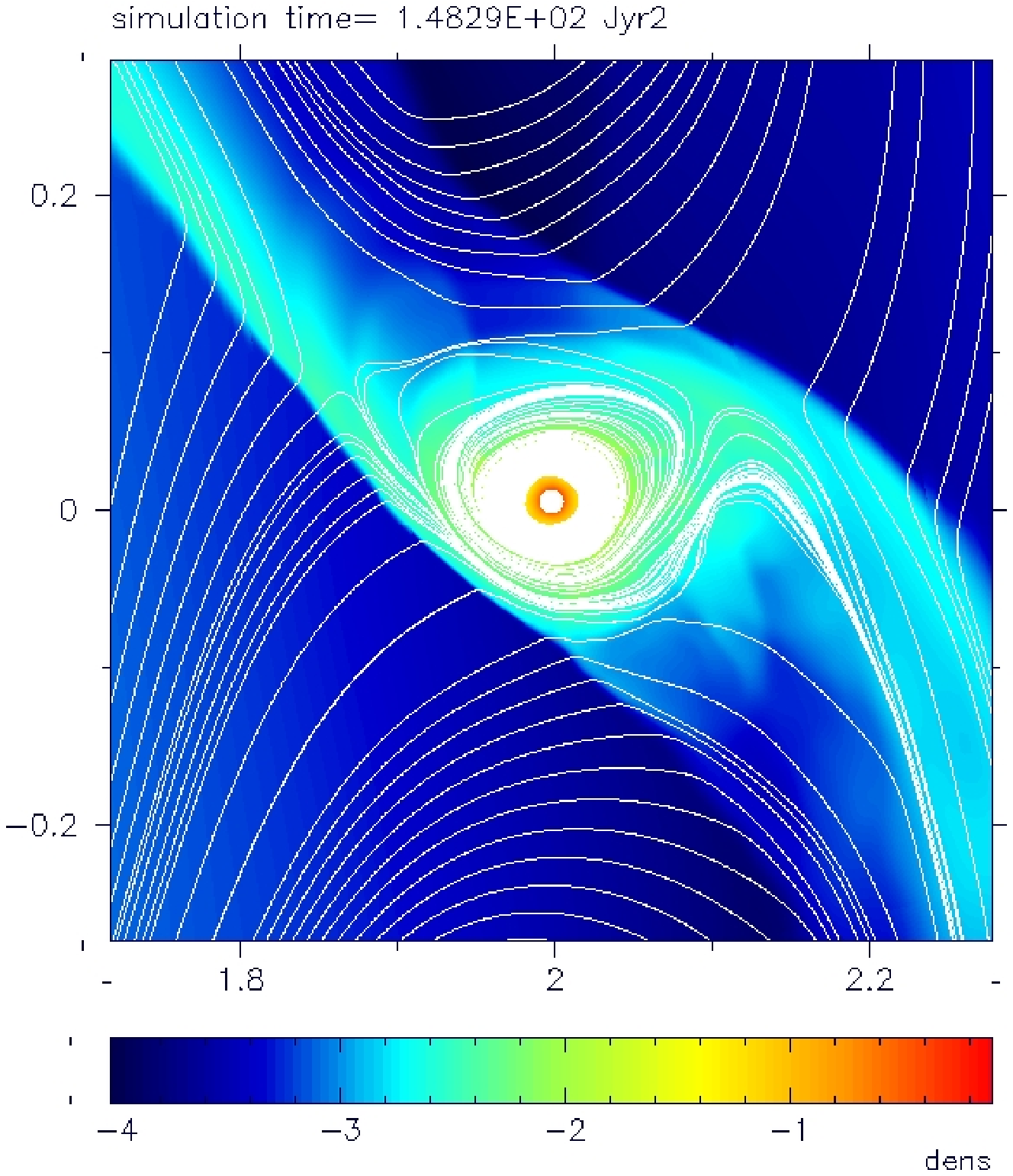}
\caption{Surface density and the flow lines in the planet's vicinity
for an inward migrating Jupiter. Different plots correspond to
different temperature profiles. Upper-left: EOS1 (no dependence on
the planet's position). Upper-right, lower-left and lower-right: EOS2
with $h_\rmn{p}$ is equal $0.2$ , $0.3$ and $0.4$ respectively. The
plotted domain is a square region of the size of $4 R_\rmn{H}$.  The
colour scale is logarithmic.}
\label{f4_flow_rl_temp_c}
\end{figure*}

\begin{figure*}
\includegraphics[width=84mm]{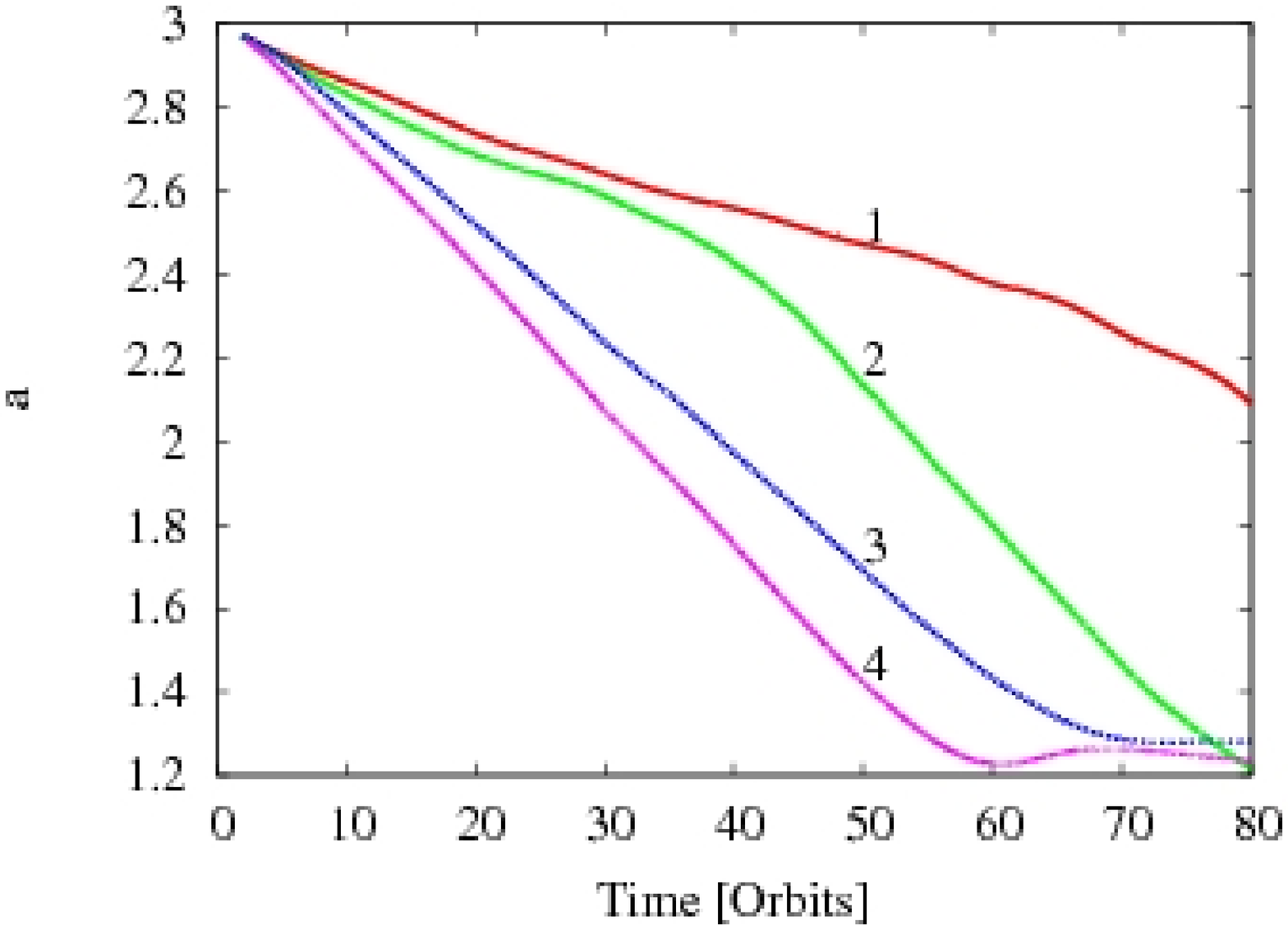}
\includegraphics[width=84mm]{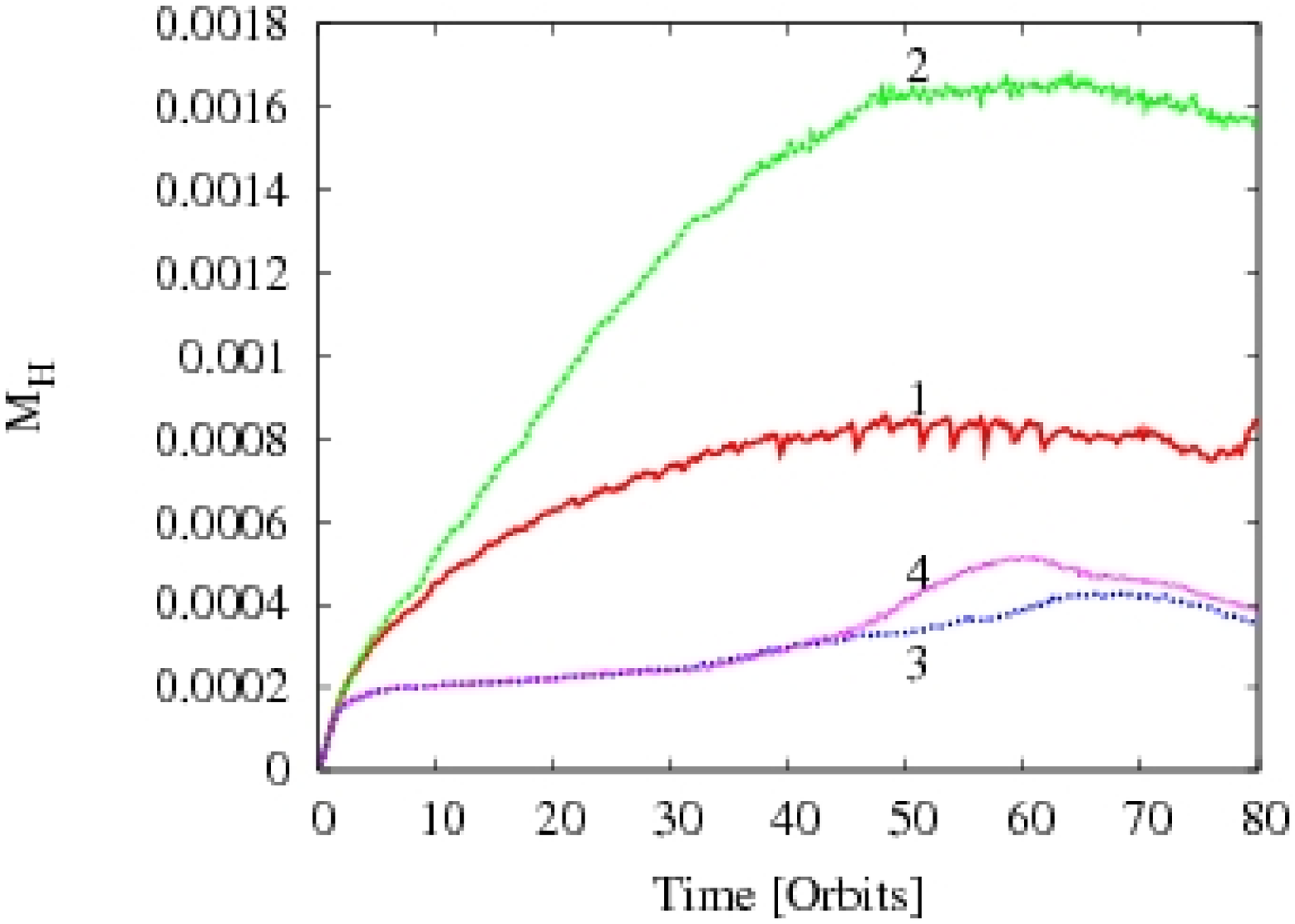}
\caption{The planet's orbital evolution for four discussed models: the
simplest model without any corrections (curve 1), the model with
correction of the gas acceleration (curve 2), the model with changed
temperature profile (curve 3) and the final model with with the
correction to the gas acceleration applied and the temperature rising
in the planet vicinity (curve 4). The left panel shows the time
evolution of the planet's semi-major axis $a$, the right panel the
mass of the gas inside a Hill sphere $M_H$. The planet mass is
$M_\rmn{P} = 0.001$.}
\label{f9_ac}
\end{figure*}


\subsection{Temperature profile in the circumplanetary disc}

\label{conv_temp_prof}

As explained in Sect.~\ref{sec_eqn_of_state}, the use of EOS1 imposes
a very thin circumplanetary disc. In such a thin disc waves can
freely propagate and the shocks in the circumplanetary disc can end
very close to the planet (see Fig.~\ref{f4_srl_cut} curve 1). Such a strong
shocks modify the flow inside the Roche lobe, redirecting the gas to
the planet's proximity. In the case of a planet on a constant orbit,
this configuration does not cause any serious problems, since the
co-orbital flow is weak and the gas inflow into the circumplanetary
disc is limited. This changes for a migrating planet. In this case
the inflow is potentially much larger since the planet moves into previously
undisturbed disk regions. In fact, mass is often seen accumulating
in the circumplanetary disc.

The flow structure of the gas in the planet's proximity for the rapid
inward migrating planet is presented in Fig.~\ref{f4_flow_rl}. On the
plot we indicate the four important regions: the inner disc I, the
outer disc II, co-orbital region III and the circumplanetary disc IV. 
For slow (type II) migration the structure of the shocks in the
circumplanetary disc and the flow pattern in the Hill sphere is
almost symmetric, but for the rapidly migrating planet the flow symmetry
is broken. The spiral shocks in the
circumplanetary disc are very time dependent and are frequently
replaced by a one-arm spiral. It is caused by a strong mass inflow 
into the circumplanetary disc. The horseshoe region shrinks to a single
tadpole-like region and a strong co-orbital flow (marked V and VI)
develops. The co-orbital flow interacts with the bow shocks and
finally is divided into a flow transferring matter from the inner disc
into the outer disc and a stream entering the Roche lobe and
accumulating in the circumplanetary disc.

Due to the strong co-orbital flow a significant amount of material is
moved into the Roche lobe and starts interacting with the shocks
originating in the circumplanetary disc. Since the disc aspect ratio
$h$ decreases with decreasing distance to the planet, the shocks
become stronger and the material moves along the shock directly to the
central density spike (see curve 1 on Fig.~\ref{f4_srl_cut}). The
radial size of this spike is given by gravitational softening
$r_\rmn{soft}$, since at this position $h$ grows and the shocks
disappear. Inside this region (about $0.6 r_\rmn{soft}$) gas moves on
circular orbits. When using EOS1 any amount of material can be added
to this disc, since we neglect self-gravity and the gas does not heat
up during compression. As a result the mass of the circumplanetary
disc is only limited by the grid resolution and the value of
$r_\rmn{soft}$.

This is the cause of the failing numerical convergence for the runs
with $r_\rmn{env}=0.5 R_\rmn{H}$ described in the previous
section. From Fig.~\ref{f2_ac2} we see that the
migration rate for the simulation with 4 refinement levels (curve 3)
increases after the inflow into the circumplanetary disk stops at
about $40$ orbits. This happens because gas moved from the inner to
the outer disc exchanges with the planet twice as much angular
momentum as the `accreted' gas.

To address this problem we replace EOS1 with EOS2. The latter maintains
a constant disc aspect ratio in the circumplanetary disc and keeps $h$
independent of the gravitational softening. To test the effects of
this different equation of state, we performed simulations with
different aspect ratios of the circumplanetary disc ($h_\rmn{p}$ equal
0.2, 0.3 and 0.4) at different resolutions (3, 4 and 5 refinement
levels), while neglecting our self-gravity correction ($r_\rmn{env} =
0$) and keeping the planet mass constant at $M_\rmn{P}$ during the
whole simulation.

We find that the model with $h_\rmn{p}=0.2$ gives no improvement over
the EOS1 results, the case with $h_\rmn{p}=0.3$ gives some, but only
for $h_\rmn{p}=0.4$ numerical convergence for 5 refinement levels was
achieved. These results can be understood by studying the density
profile near the planet (Fig.~\ref{f4_srl_cut}) and density and flow
patterns (Fig.~\ref{f4_flow_rl_temp_c}).  The $h_\rmn{p}=0.2$ case
gives a profile very similar to EOS1.  The model with $h_\rmn{p}=0.3$
results in a somewhat smoother profile, but only for $h_\rmn{p}=0.4$
the density profile is smooth enough for the shocks to become
unimportant, and the gas inflow into the circumplanetary disc is
stopped, resulting in a lower mass inside the Roche lobe. The planet's
orbital evolution and the mass of the Hill sphere content for the
$h_\rmn{p}=0.4$ simulation at different refinement levels are shown in
Fig.~\ref{f7_af4}.

These results seem to indicate that the change of temperature profile
alone is sufficient for achieving numerical convergence, even if we do
not apply any corrections for self-gravity. The reason for this is the
smoother density distribution and the lower mass of the gas in the
circumplanetary disc. The `artificial inertia' effects are
particularly strong when the mass of the circumplanetary disk is high,
and its density profile strongly peaked. In this case any small
discrepancy between the disc centre and planet position results in
strong, high frequency oscillations of all orbit
parameters.\footnote{This oscillations are not visible on the plots,
since we removed them by averaging over $5$ orbits.}. 

However, even for the changed temperature profile, some `artificial
inertia' effects remain. To illustrate this we compare the planet's
orbital evolution for four different models in Fig.~\ref{f9_ac}. Model
1 is the simplest model (without any self-gravity correction and with
EOS1), model 2 uses the self-gravity correction ($r_\rmn{env} = 0.5
R_\rmn{H}$), model 3 employs EOS2 ($h_\rmn{p} = 0.4$) but no
self-gravity correction, and model 4 applies both ($r_\rmn{env} = 0.5
R_\rmn{H}$ and $h_\rmn{p} = 0.4$). We see that models 2, 3 and 4 show
a higher migration rate than model 1. For model 2 there is a visible
kink at $45$ orbits, which is the moment when the limiting mass of the
circumplanetary disc is reached. This kink is invisible for model 4,
since the `accreted mass' is an order of magnitude smaller, and this
mass is already reached after the few first orbits. Comparing models 3
and 4 we see that the latter has a larger migration rate than the
third model, implying that the change of the temperature profile does
not remove the effects of `artificial inertia' completely. We hence
should consider both mechanisms to be equally important.

\begin{figure*}
\includegraphics[width=84mm]{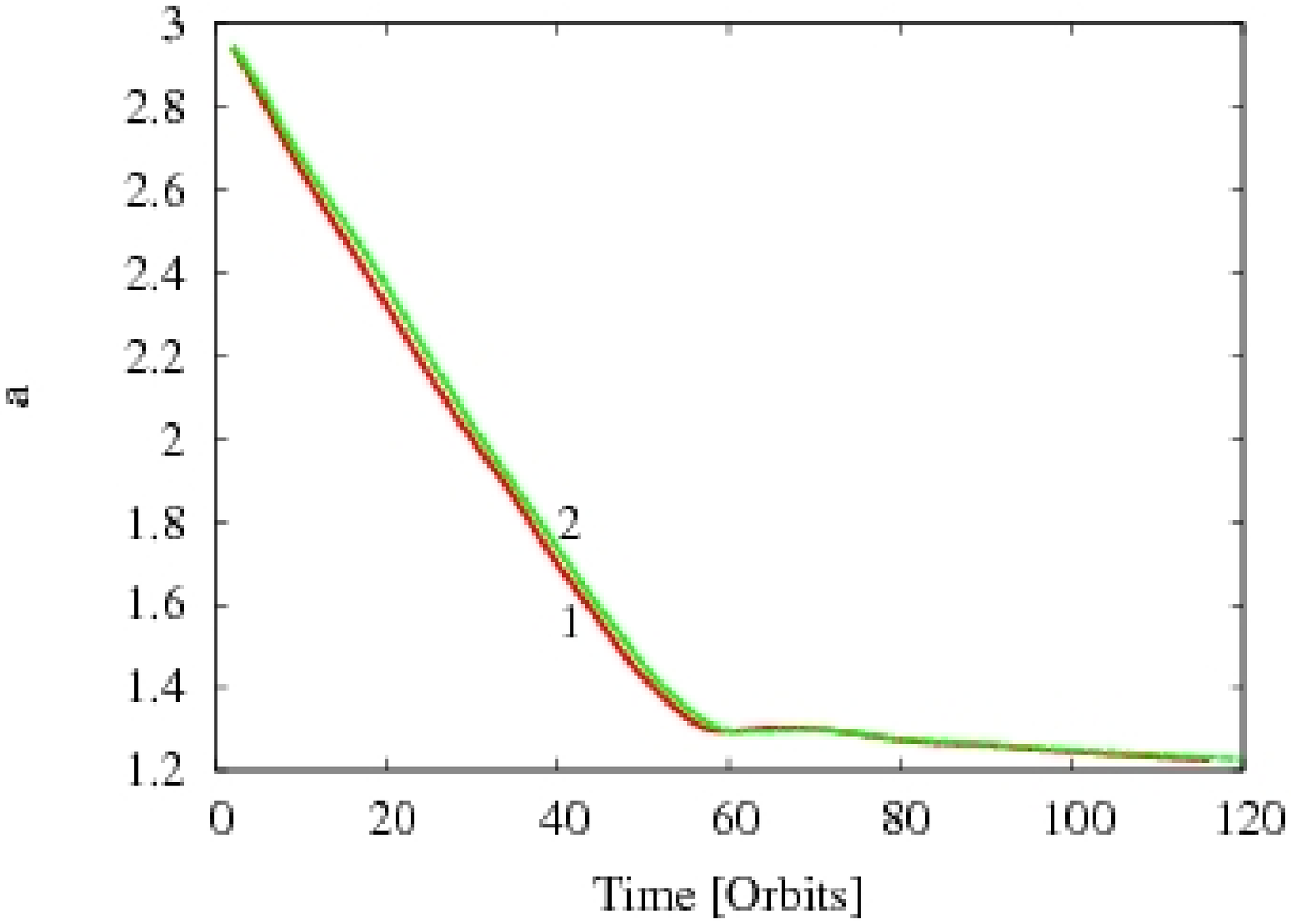}
\includegraphics[width=84mm]{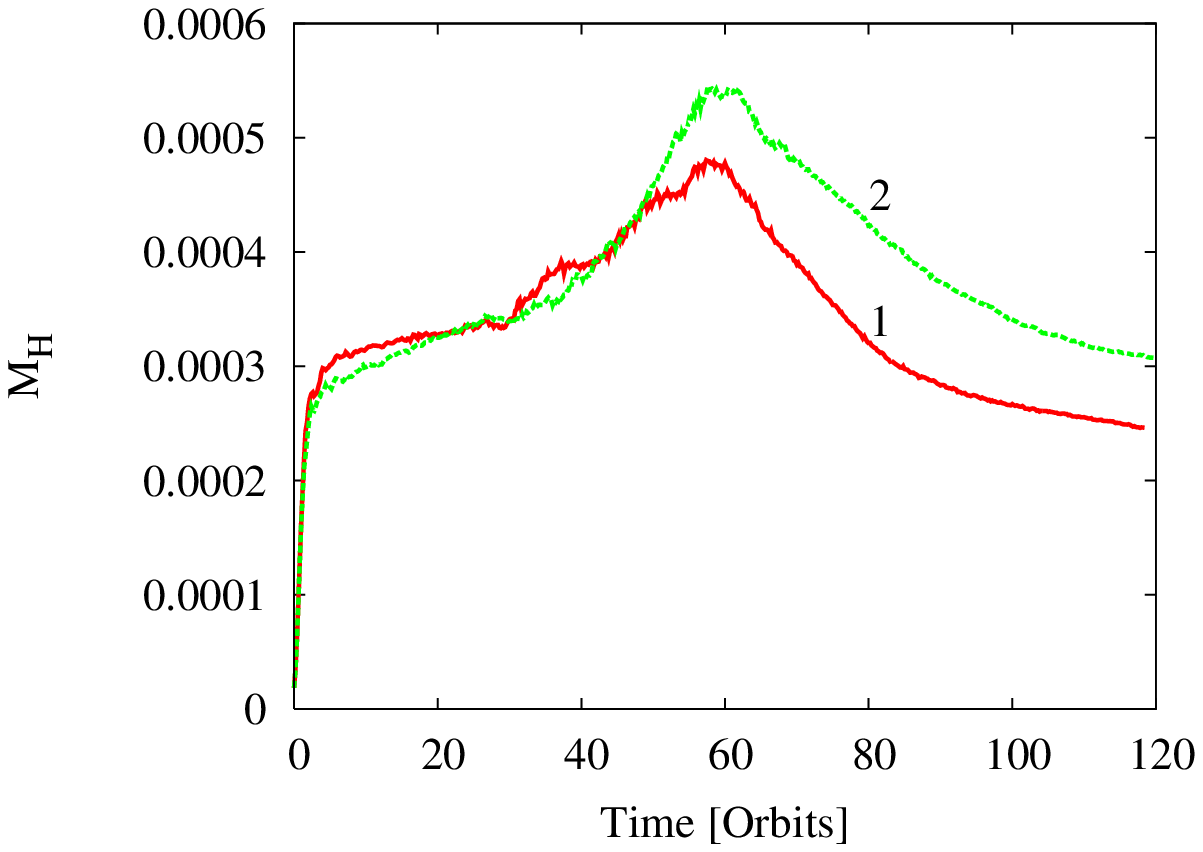}
\caption{The planet orbital evolution for different smoothing lengths
of the planet's gravitational field for $h_\rmn{p} = 0.4$. 
The time evolution of the planet's semi-major axis $a$ (left panel) 
and the mass of the gas inside a Hill sphere $M_H$ (right panel) are 
plotted. The planet mass $M_\rmn{P} = 0.001$. Curves 1 and 2 correspond to
$r_\rmn{soft}$ equal $0.0208$ and $0.3 R_\rmn{H}$ respectively.}
\label{f15_a_soft_4l}
\end{figure*}

\begin{figure}
\includegraphics[width=84mm]{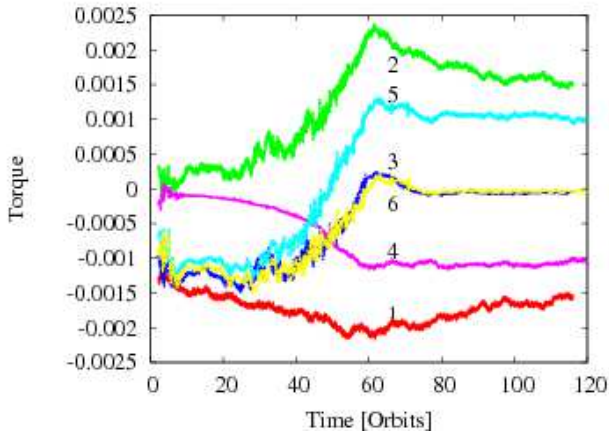}
\caption{The torque exerted by the gas on the planet for simulations with 
different smoothing length of the planet gravitational field $r_\rmn{soft}$ 
equal $0.0208$ (curves 1, 2 and 3) and $0.3 R_\rmn{H}$ (curves 4, 5 and 6).
The torque was calculated separately for the interior of the planet's gravitational
softening $r_\rmn{p} < r_\rmn{soft}$ (curves 1 and 4) and the rest of the Roche sphere 
$r_\rmn{soft} < r_\rmn{p} < R_\rmn{H}$ (curves 2 and 5), where $r_\rmn{p}$ is the d
istance to the planet. The curves 3 and 6 give the torque from the whole Hill sphere 
$\Gamma_\rmn{RL}$ (the sum of the curves 1, 2 and 4, 5 respectively).}
\label{f15_trq_soft_4l}
\end{figure}

\begin{figure*}
\includegraphics[width=84mm]{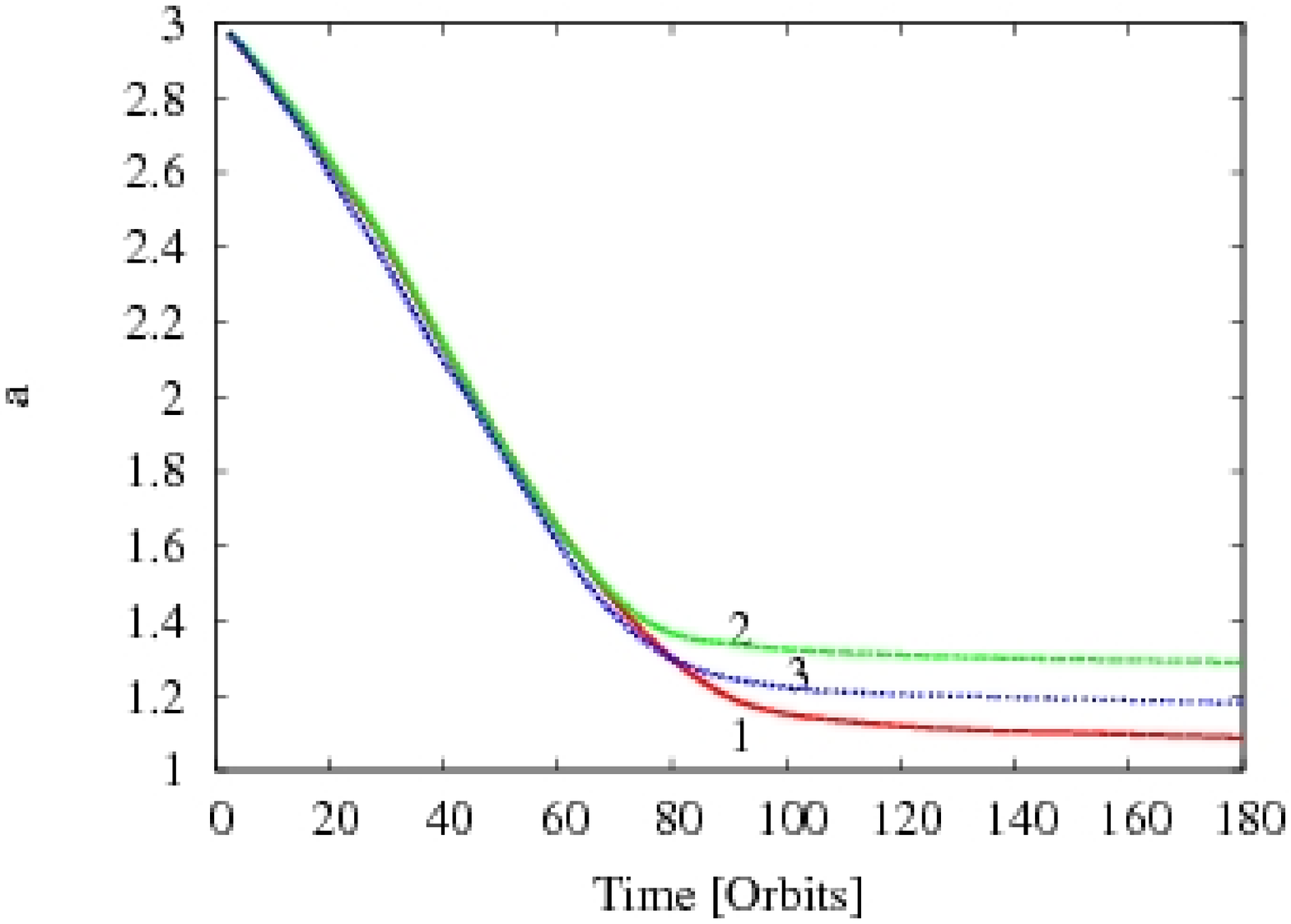}
\includegraphics[width=84mm]{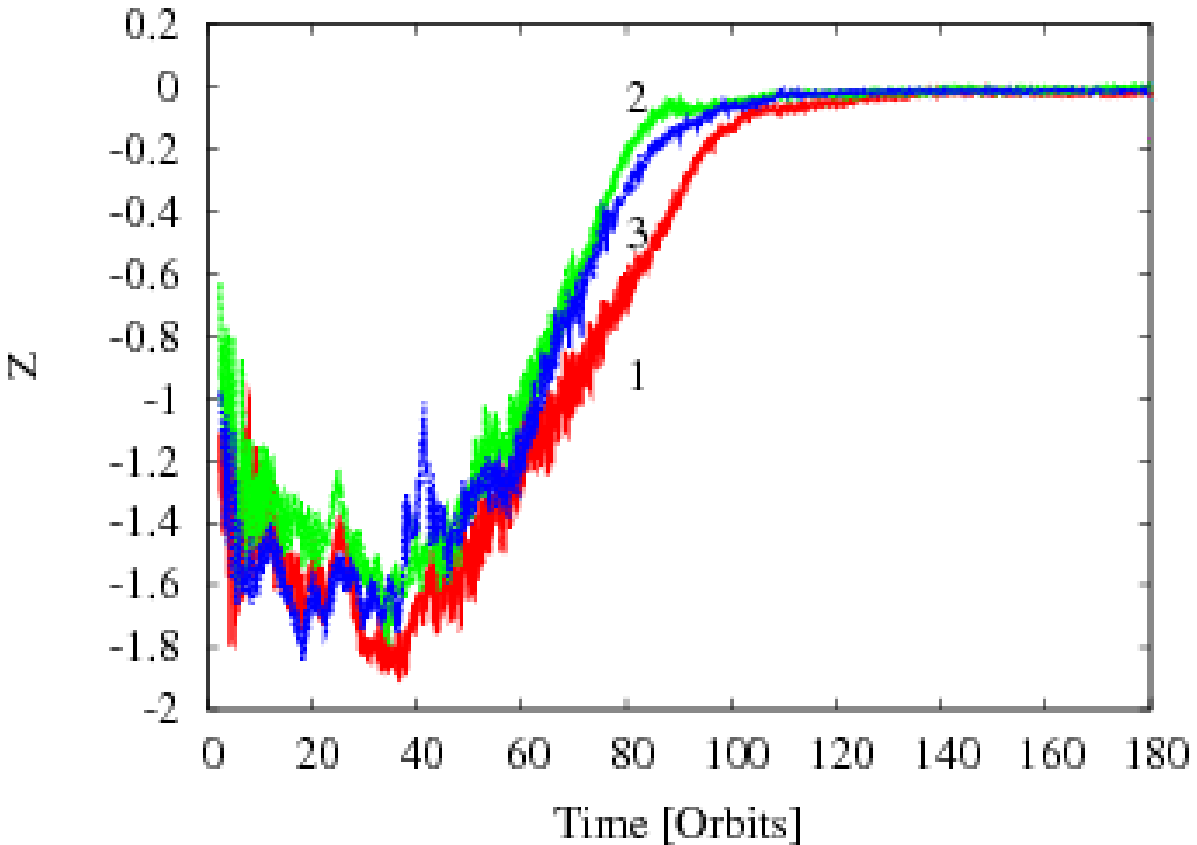}
\includegraphics[width=84mm]{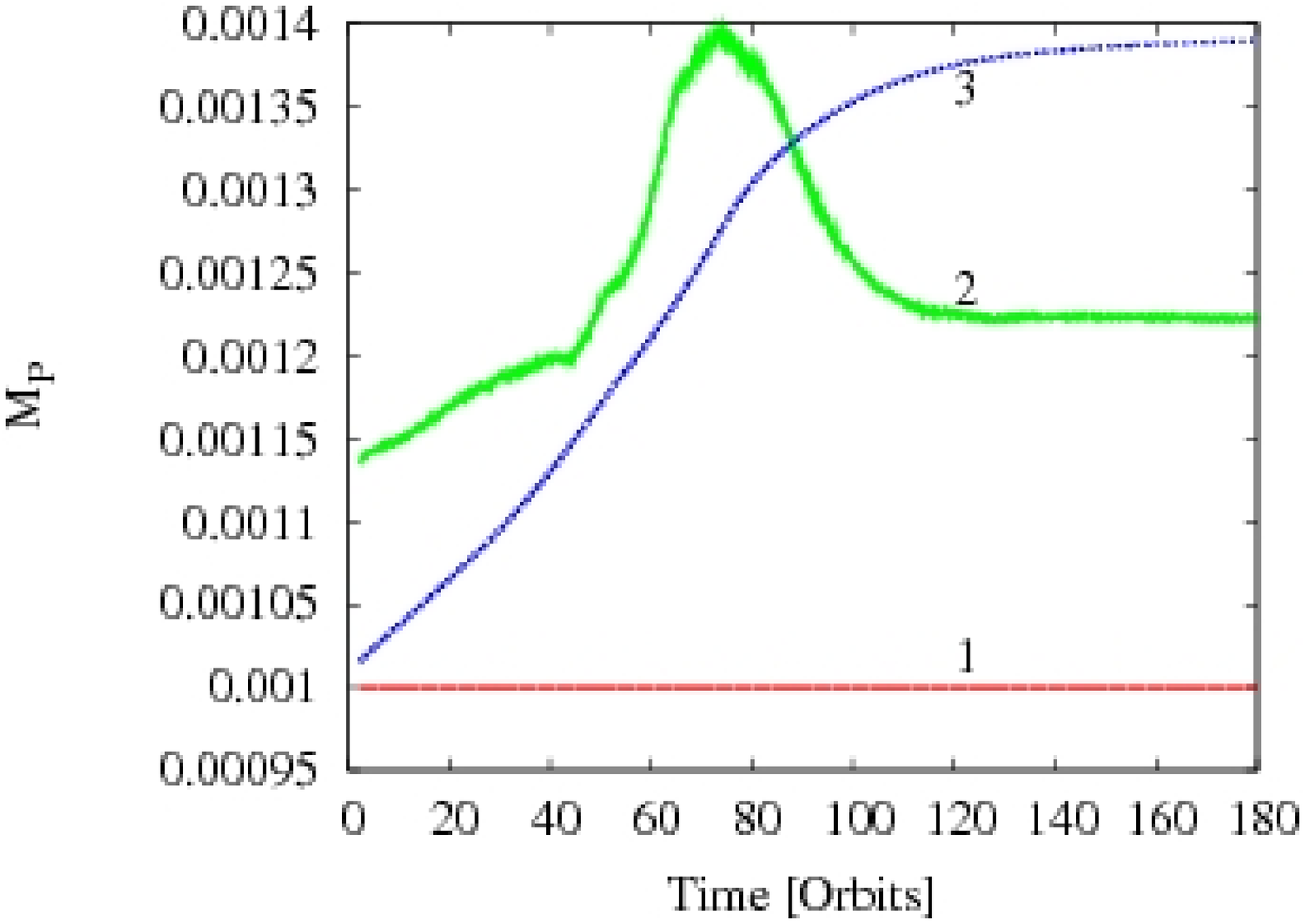}
\includegraphics[width=84mm]{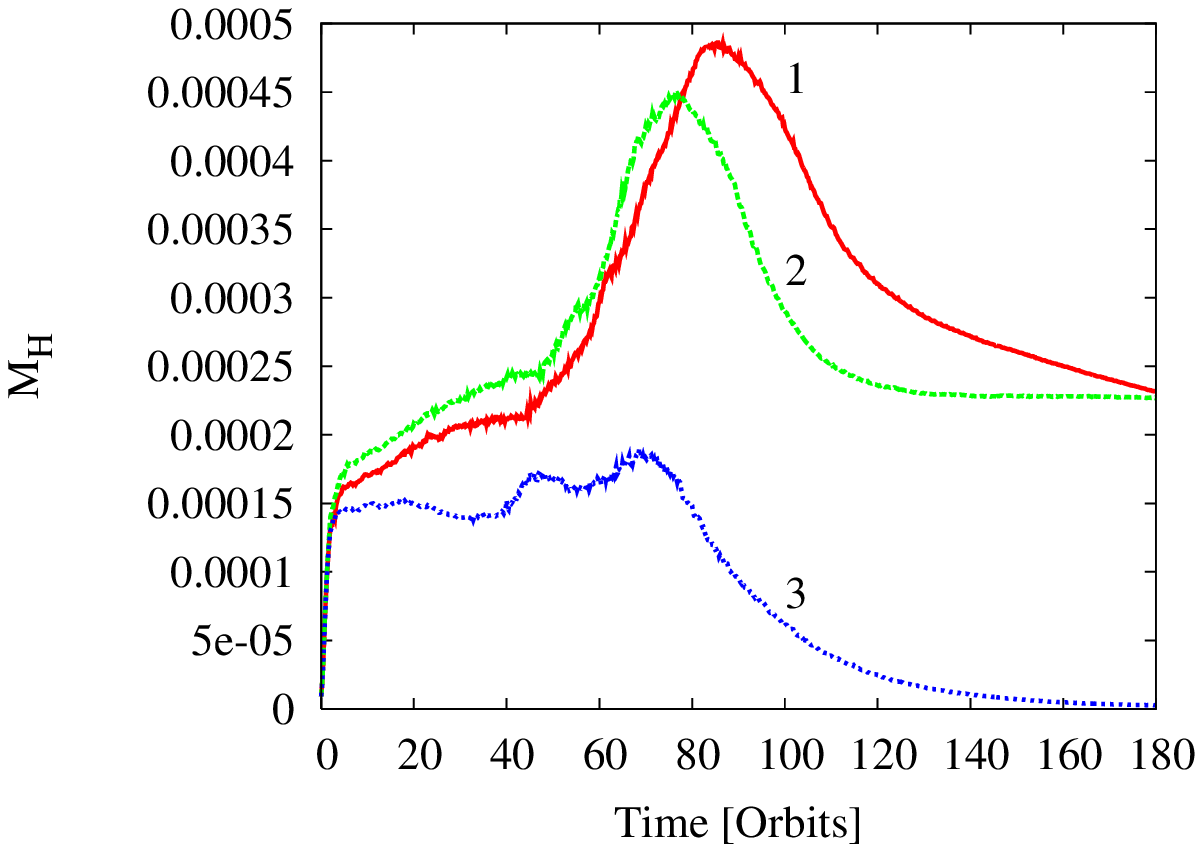}
\caption{ Results of the simulations for the varying planet's
mass and $h_\rmn{p}=0.4$. Curve 1 corresponds to the model with 
the constant planet's mass $M_\rmn{P} = 0.001$. Curve 2 shows the 
simulation with the effective  mass of the planet augmented with the 
mass within the planet's
smoothing length ($\widetilde M_\rmn{P}$). Curve 3 presents the model
with an accreting planet. Upper left and upper right panels show the
orbital evolution and the non-dimensional migration rate $Z$
respectively. The planet mass and the mass of the gas inside a Hill
sphere are presented on the lower left and lower right panels. During
the first $70$ orbits (fast migration limit $|Z|>1$) all systems
evolve almost exactly the same way. Although in the fast migration
regime the migration rate $\dot a$ is identical in simulations 1
and 2 and the differences for the accreting model are small, $Z$
differs slightly, since $\dot a_\rmn{f}$ depends on the planet
mass. The simulations start to differ after reaching the slow
migration limit.}
\label{fmin_d_planet_mass}
\end{figure*}

\section{Torque calculation and effective planet mass}
\label{sect_tor_mass}

In the previous section we discussed the modifications of the disc 
model that allow to achieve the numerical convergence. In this chapter 
we will concentrate on the calculation of the torque inside the Hill 
sphere, and the models with the varying planet mass. Especially we 
will focus on the dependence of the migration on the choice of $M_\rmn{P}^*$. 
We will discuss it for the inward and outward migration case separately.

\subsection{Inward migration case}

\subsubsection{Gravitational softening}

\label{conv_grav_soft}

When including all material inside the Roche lobe in the calculation
of the torque, the value for the gravitational softening
$r_\rmn{soft}$ can play an important role, even though increasing the
value of the circumplanetary disc aspect ratio $h_\rmn{p}$ should make
$\dot a$ less dependent on $r_\rmn{soft}$. Indeed we find that for
$h_\rmn{p} \geq 0.4$ the migration behaviour does not depend on
$r_\rmn{soft}$.  The results of the runs with $h_\rmn{p} =0.4$ and
$r_\rmn{soft}$ equal $0.0208$ and $0.3 R_\rmn{H}$ are presented in
Fig.~\ref{f15_a_soft_4l}. In the first case
$r_\rmn{soft}$ is constant during the simulation and its value in
$R_\rmn{H}$ units is ranging from $0.1 R_\rmn{H}$ (initial value) up
to about $0.25 R_\rmn{H}$.  In both simulations we used $M^*_\rmn{P} =
M_\rmn{P}$. The first and the second plot show the planet orbital
evolution and the mass of the gas inside a Hill sphere. The orbital
evolution is independent on the size of smoothing lengths of the
planet's gravitational field, however a larger $r_\rmn{soft}$ allows the
planet to accumulate larger amounts of material.

Above we have argued against excluding any part of the disc in the
torque calculation. The results in this section allow this to be
illustrated better.  Figure~\ref{f15_trq_soft_4l} presents the torque
exerted on the planet by the gas contained within the Hill sphere. Curves
3 and 6 show the torque from the entire Hill sphere
$\Gamma_\rmn{RL}$. We divided this torque into two parts, namely the
contribution from $r_\rmn{p} < r_\rmn{soft}$ ($\Gamma_\rmn{soft}$,
curves 1 and 4) and the contribution from $r_\rmn{p} > r_\rmn{soft}$
($\Gamma_\rmn{out}$, curves 2 and 5), where $r_\rmn{p}$ is the
distance to the planet. The two set of curves (1, 2, 3) and (4, 5, 6)
correspond to $r_\rmn{soft}$ equal $0.0208$ and $0.3 R_\rmn{H}$
respectively.

$\Gamma_\rmn{RL}$ is driving the migration during the fast migration
phase (first $60$ orbits) and drops in the slow, type II like
migration phase. As we saw before, it is almost independent on
$r_\rmn{soft}$.  However, the partial torques $\Gamma_\rmn{soft}$ and
$\Gamma_\rmn{out}$ are varying strongly between the two choices for
$r_\rmn{soft}$. During the fast migration phase even the sign of
$\Gamma_\rmn{out}$ depends on $r_\rmn{soft}$. During the slow phase
the two choices for the softening/cutting radius both give a positive
$\Gamma_\rmn{out}$, but its value increases with decreasing
$r_\rmn{soft}$, consistent with what was found by
\citet{2002A&A...385..647D}.  $\Gamma_\rmn{soft}$ has a similar but
negative value and thus $\Gamma_\rmn{RL}$ is close to zero,
independent of the value of $r_\rmn{soft}$.

This illustrates the point made before: the interior of the Roche lobe
is a complicated and variable dynamical system, and it is difficult to
define a simple spherical region of radius $r_\rmn{cut}$, that
dynamically belongs to the planet and could be neglected in the torque
calculation. Any fast migration calculation would be very sensitive to
the choice of $r_\rmn{cut}$ \citep{2005MNRAS.358..316D}, and a wrong
choice can give systematic errors in the torque calculation. It is
therefore better to include all of the torques, even though this
implies dealing with the region very close to the planet.

\subsubsection{Mass accumulation in the planet's vicinity}
\label{conv_grav_mass_in}

\begin{figure*}
\includegraphics[width=84mm]{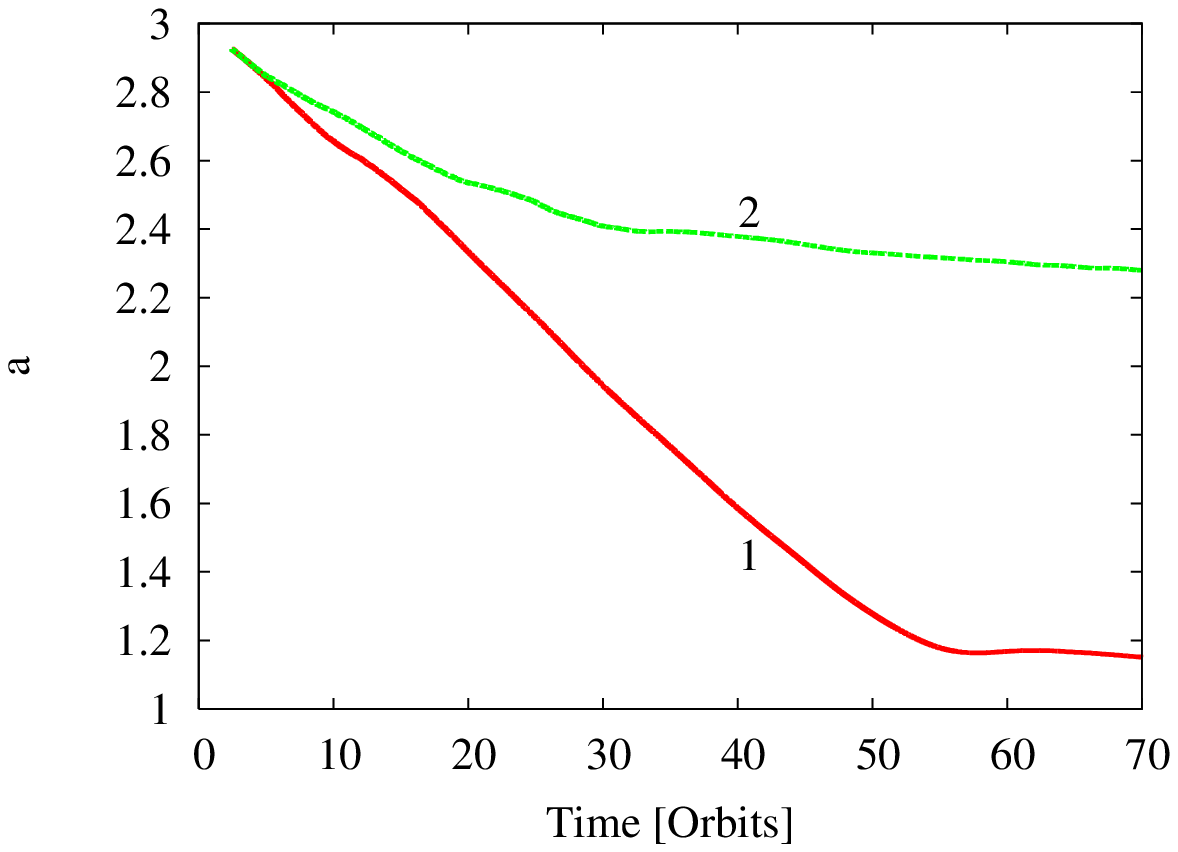}
\includegraphics[width=84mm]{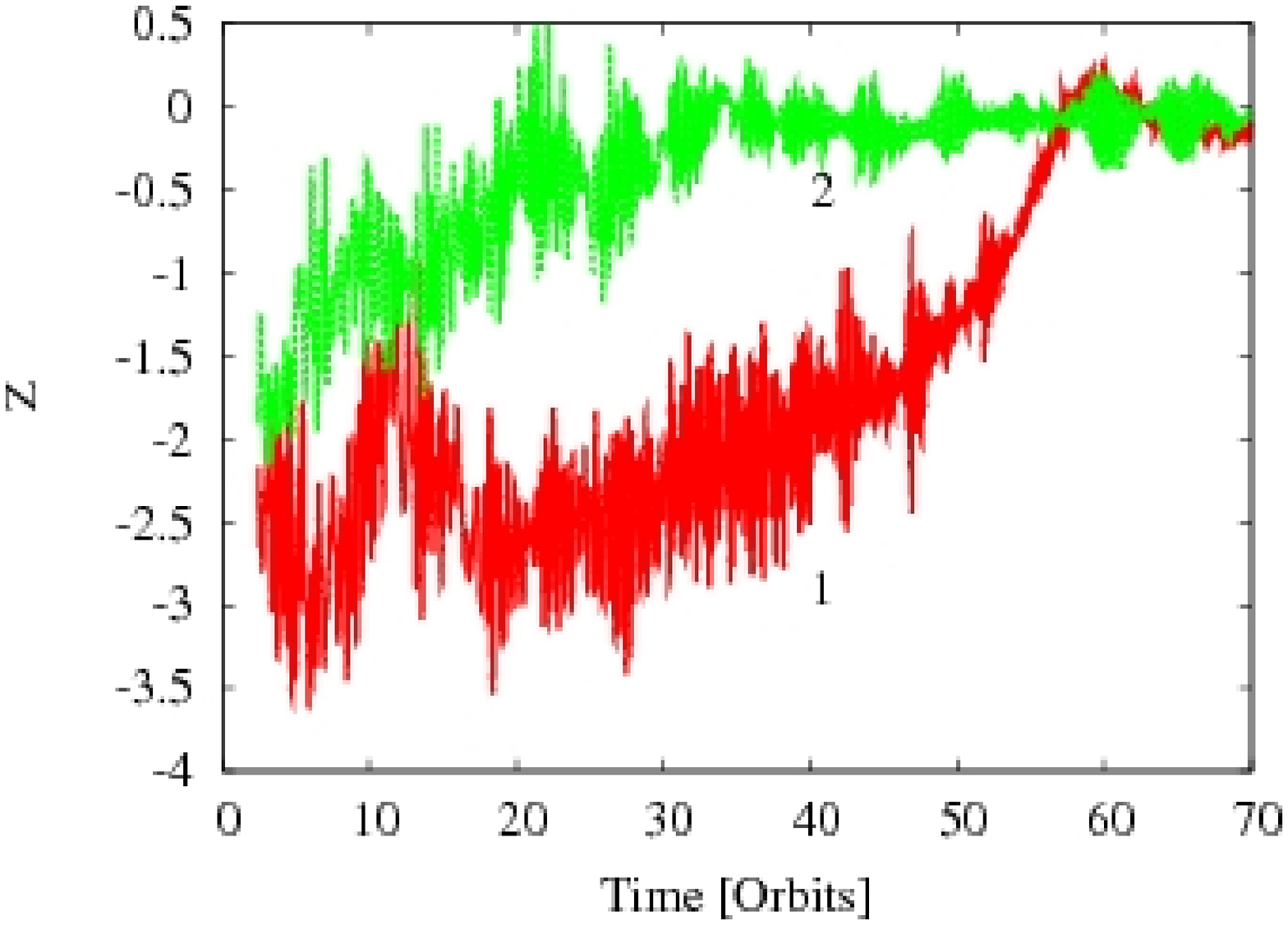}
\includegraphics[width=84mm]{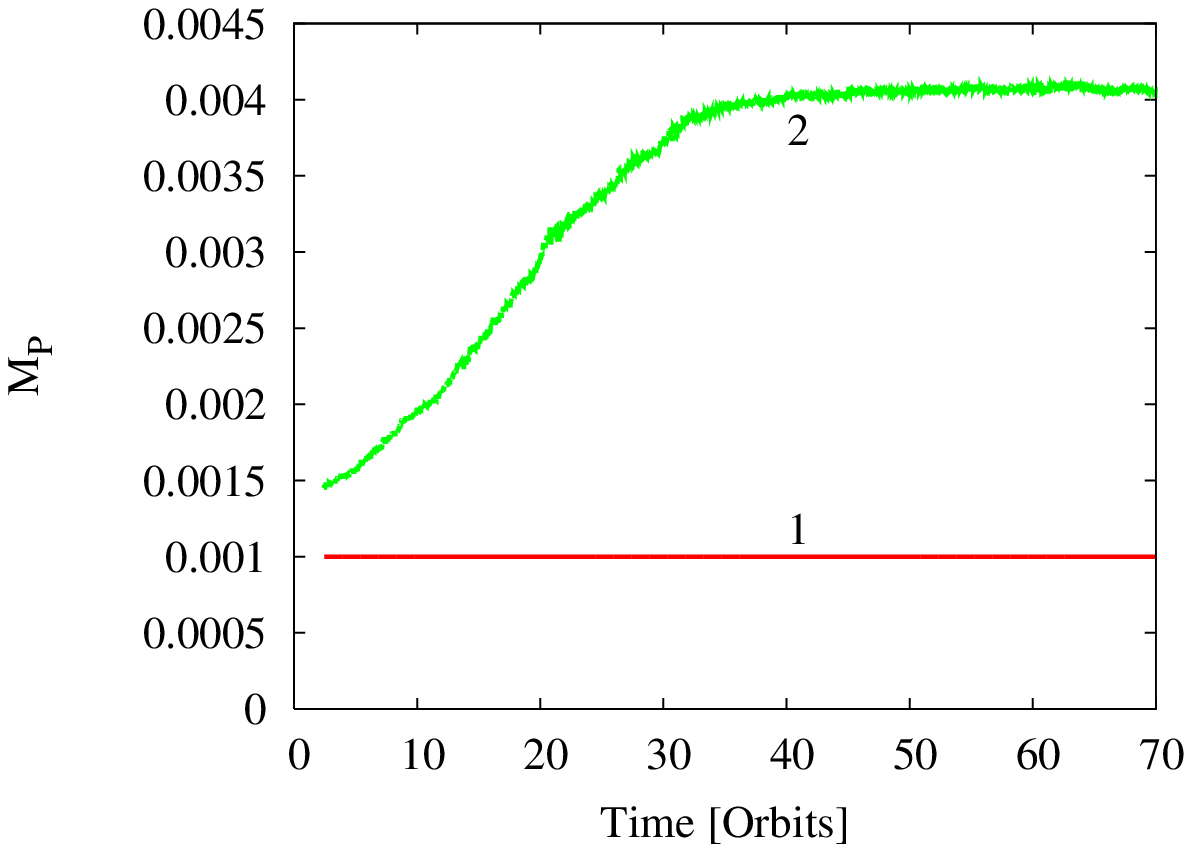}
\includegraphics[width=84mm]{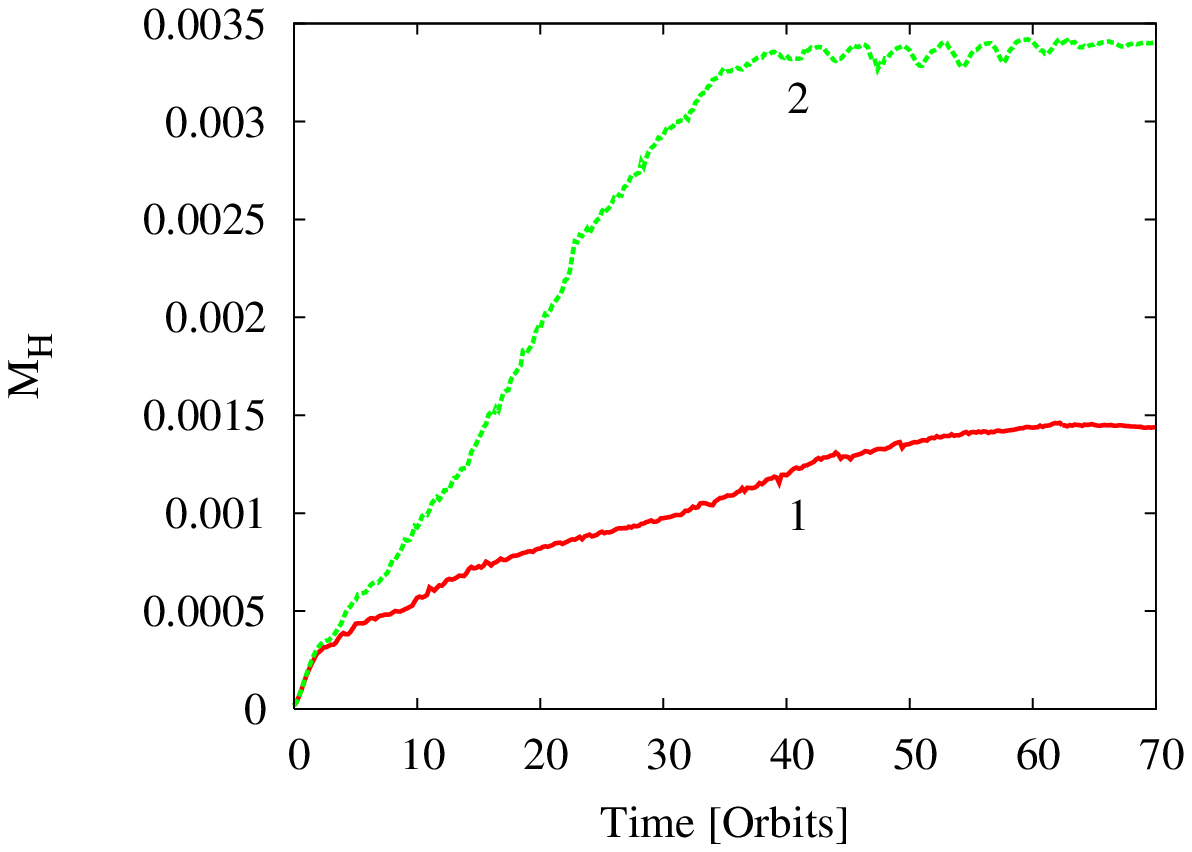}
\caption{Results of the simulations of inward migrating planet with
the varying planet's mass for $h_\rmn{p}=0.3$. Curve 1 corresponds 
to the model with the constant planet's mass $M_\rmn{P} = 0.001$. 
Curve 2 shows the simulation with the effective planet mass 
increased by the mass within the planet's smoothing length. Upper 
left and upper right panels show the orbital evolution and the 
non-dimensional migration rate $Z$ for the inward migrating planets 
respectively. The planet mass and the mass of the gas inside a Hill 
sphere are presented on the lower left and lower right panels.}
\label{fmin_d_planet_mass_hp}
\end{figure*}

Up to this point we studied models with a constant planet mass, neglecting the
gravitational interaction of the planet's envelope with the whole
disc.  However, the mass of gas accumulating in the planet's
vicinity in the case of constant $h_\rmn{p} = 0.4$ is comparable to
the initial planet mass and can influence the orbital evolution. In
this section we present models where we vary the planet's mass.

To test the dependence of the planet's orbital evolution on a varying
planet mass we performed three simulations. In the first one the
effective planet mass is constant during the whole simulation
$M^*_\rmn{P}=M_\rmn{P} = 0.001$, similar to the models
presented above. In the second model the planet mass was replaced with
all of the mass within the smoothing length $r_\rmn{soft}$ around the
planet ($M^*_\rmn{P}=\widetilde M_\rmn{P}$). In the last model we studied the
planet growth through gas accretion, i.e.\ removing gas from
the planet's environment and adding its mass and momentum to the
planet (see Sect~\ref{desc_mod_accr}).

The results are presented in Fig.~\ref{fmin_d_planet_mass}.  Curves 1
and 2 correspond to the constant mass and the $\widetilde M_\rmn{P}$
case respectively.  Curve 3 presents the accreting model. The upper left
and upper right panels show the orbital evolution and the
non-dimensional migration rate $Z$ (see below). The planet mass and
the mass of the gas inside a Hill sphere are presented on the lower
left and lower right panels.

The non-dimensional migration rate $Z$ is defined as the ratio of the
migration rate ${\dot a}$ and the so-called fast migration
speed ${\dot a_\rmn{f}}$ 
\begin{equation}
Z = {{\dot a} \over {\dot a_\rmn{f}}},
\end{equation}
where $\dot a_\rmn{f}$ is given by the ratio of the
half width of the horseshoe region $x_\rmn{s}$ and the libration
time-scale $T_\rmn{lib}$. In a Keplerian disc $\dot a_\rmn{f}$ equals
(\citealt{PaperIV}, \citealt{2007prpl.conf..655P})
\begin{equation}
\dot a_\rmn{f} = {{3 x_\rmn{s}^2 \Omega} \over {8 \pi a}},
\end{equation}
where $x_\rmn{s}$ is estimated to be about $2.5$ Hill sphere
radii. We can divide the type III migration into a fast, $|Z|>1$, and a slow,
$|Z|<1$, migration regime. Type II-like migration phase
corresponds to $|Z| \ll 1$. The use of $Z$ will be discussed more
fully in Paper II.

During the fast migration phase (lasting for about $70$ orbits)
the planet's orbital evolution is almost independent of the planet's
mass and the mass of the circumplanetary disc. All three models show a
similar evolution even though the planet's effective mass can
differ up to $40\%$ and the mass of the circumplanetary disc can
differ up to $100\%$. The amount of the gas flowing into the Hill
sphere (for the model with accretion the sum of the Hill sphere
content and the increase of the planet's mass) is similar too. This
means that the inflow into the circumplanetary disc is a dynamical
effect that mostly depends on the value of $Z$ and the initial disc
surface density, and only weakly depends on the gas dynamics deep in
the Roche lobe. On the other hand, it should be very sensitive to the
temperature profile at the boundary of the Roche lobe. Comparing the
lower left and the lower right panels of Fig.~\ref{fmin_d_planet_mass}
we can see that most of the mass within the Roche sphere is contained
within the planet's smoothing length.

The differences between the three cases show up in the slow migration
phase, where the first and the second model lose gas from the Hill
sphere. The simulation with constant $M_\rmn{P}$ keeps losing mass
from the circumstellar disc during that entire phase, whereas in the
second simulation the amount of mass within the Hill sphere converges
to $1.22 M_\rmn{P}$. In the accreting model the amount of gas
available in the planet's vicinity and the pressure gradient are too
small to cause mass loss from the circumplanetary disc. Instead the
gas orbiting the planet is accreted, and in the stage of a gap
creation (when the strong mass inflow into the circumplanetary disc
stops) the amount of the mass inside the Hill sphere quickly
decreases. At the end of the simulation the planet has reached a mass
of $1.4 M_\rmn{P}$.

The differences in the slow migration regime ($|Z|<1$) lead to
different final positions in the three cases. The reason is that
$\dot a_\rmn{f}$ depends on the planet mass. The case of a constant
planet mass makes the transition to the slow phase latest, and thus
achieves the smallest orbit at $a=1.1$. The difference between the
$\widetilde M_\rmn{P}$ case and the accreting planet is due to the
fact that the former loses mass from the Hill sphere. This influences
the migration during the end of the slow migration phase (see
the difference between curves 2 and 3 between $70$ and $90$ orbits in
the upper right panel of Fig.~\ref{fmin_d_planet_mass}) and allows the
accreting planet to migrate further ($a=1.2$) than the planet in the
model with the effective mass increased ($a=1.3$).

As we can see, the orbital evolution differs between the fast ($|Z|>1$)
and the slow  ($|Z|<1$) migration limit. During the fast
migration under constant $h_\rmn{p}$, we can neglect details of the
gas evolution deep in the circumplanetary disc, since the rate of the
mass accumulation in the planet's proximity is relatively low and the
migration is driven by the outer part of the Roche lobe. However, we
should keep in mind that a variation of $h_\rmn{p}$ can modify the
flow in the whole Roche lobe and thus can influence migration. The
gas evolution in the circumplanetary disc becomes important for $|Z|
\approx 1$, when the planet slows down and starts to open a gap, and
the rate of the mass accumulation in the planet's proximity increases.

To test this last dependence we performed simulations with 
constant planet mass (case 1 above) and with the planet's effective 
mass increased by the mass within the planet's smoothing length (case
2 above) for $h_\rmn{p}$ equal $0.3$, $0.5$ and $0.6$. The results for
$h_\rmn{p} = 0.3$ are presented in Fig.~\ref{fmin_d_planet_mass_hp}
(curves 1 and 2 for ${M_\rmn{P}}^* = 0.001$ and ${M_\rmn{P}}^* =
\widetilde M_\rmn{P}$ respectively).  The upper left and upper right
panels show the orbital evolution and the non-dimensional migration
rate $Z$, respectively. The planet mass and the mass of the gas inside
a Hill sphere $M_H$ are presented on the lower left and lower right
panels. We present here the results for $h_\rmn{p} = 0.3$ even though
the planet's migration is slightly dependent on $r_\rmn{soft}$ for
this value of the circumplanetary disc aspect ratio ($M_H$ grows from
$0.0015$ up to $0.0022$ for $r_\rmn{soft}$ changing form $0.3
R_\rmn{H}$ to $0.0208$, and the migration rate $\dot a$ differs by
less than $5\%$ between both models), since it illustrates the
interesting case, were the sound speed $c_\rmn{s}$ at the boundary of
the Roche lobe given by EOS2 is smaller than $c_\rmn{s}$ given by
EOS1. The model with $h_\rmn{p} = 0.4$ is the limiting case, where
both equation EOS1 and EOS2 give the same real disc aspect ratio
outside the Hill sphere (see Fig.~\ref{f2_hpl}). In the rest of the
models EOS2 gives bigger $c_\rmn{s}$ in the planet's vicinity than EOS1.

The amount of mass accumulated in the planet's vicinity 
is very sensitive to the value of $h_\rmn{p}$. Unlike the model with 
$h_\rmn{p} = 0.4$, the model with  $h_\rmn{p} = 0.3$ and 
${M_\rmn{P}}^* = M_\rmn{P}$ allows the accumulation 
in the circumplanetary disc of an amount of mass comparable to 
the planet mass (about $50\%$ of $M_\rmn{P}$ after $10$ orbits). In the 
second simulation (${M_\rmn{P}}^* = \widetilde M_\rmn{P}$) this mass 
grows faster, since the Hill radius grows with $\widetilde M_\rmn{P}$ 
and the non-dimensional migration rate $|Z|$ decreases below 1, allowing 
a speed up of the accumulation (see Fig.~\ref{fmin_d_planet_mass}). 
Finally $\widetilde M_\rmn{P}$ reaches $4$ Jupiter masses and the planet 
migrates in the slow migration limit, stopping in the type II like migration 
at about $a=2.2$. As we can see the planet's migration is sensitive to the 
 rapid increase of $\widetilde M_\rmn{P}$. The limiting case $h_\rmn{p} = 0.4$ 
was already discussed and shows the different behaviour in the slow and the fast 
migration limit. Increasing $h_\rmn{p}$ above $0.4$ means significant decreasing 
$M_\rmn{soft}$ and the planet's migration becomes independent of the choice 
of ${M_\rmn{P}}^*$. This allows the planet to travel faster
and further. A more detailed description will be presented in Paper~II.

\begin{figure*}
\includegraphics[width=84mm]{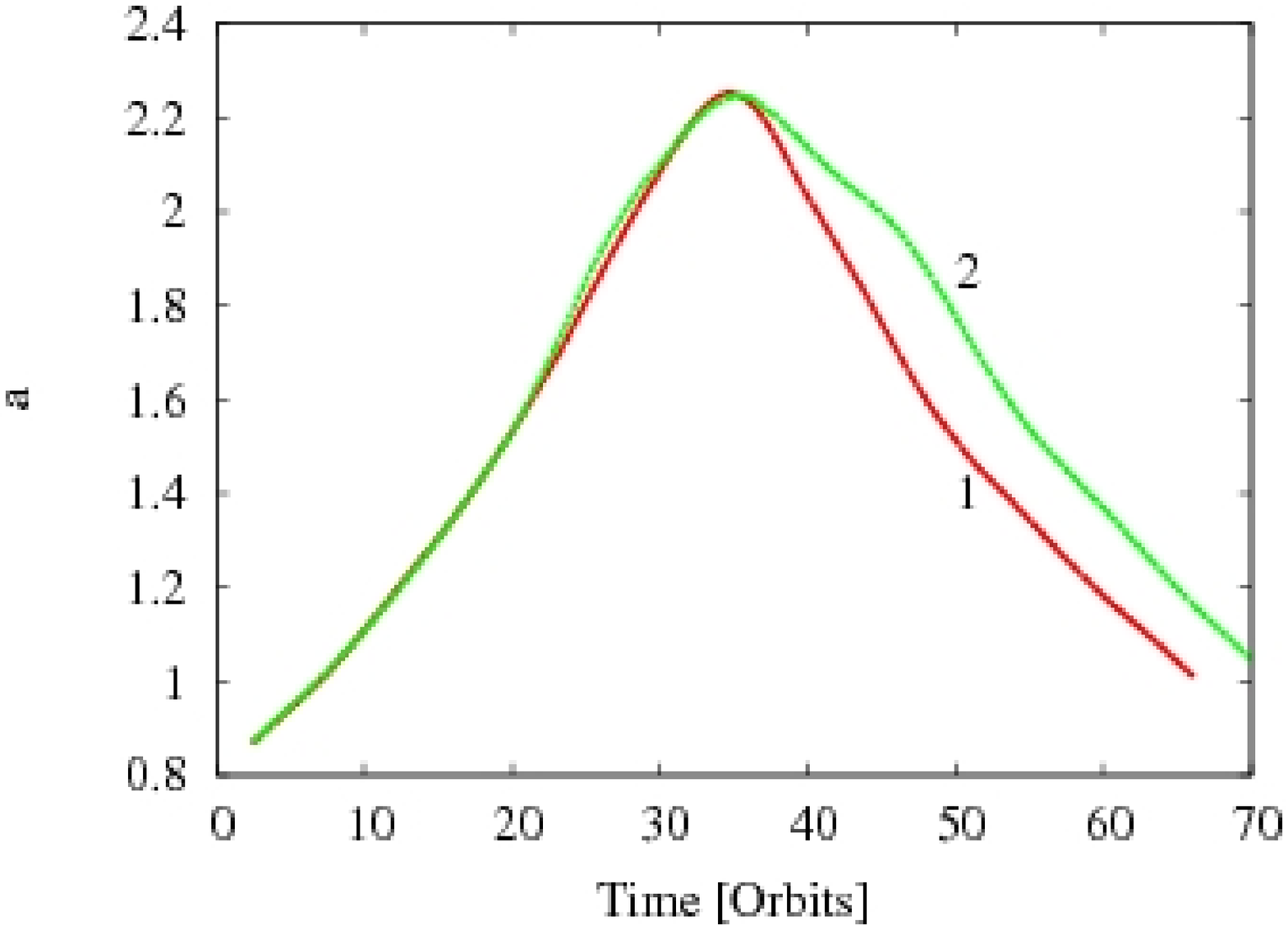}
\includegraphics[width=84mm]{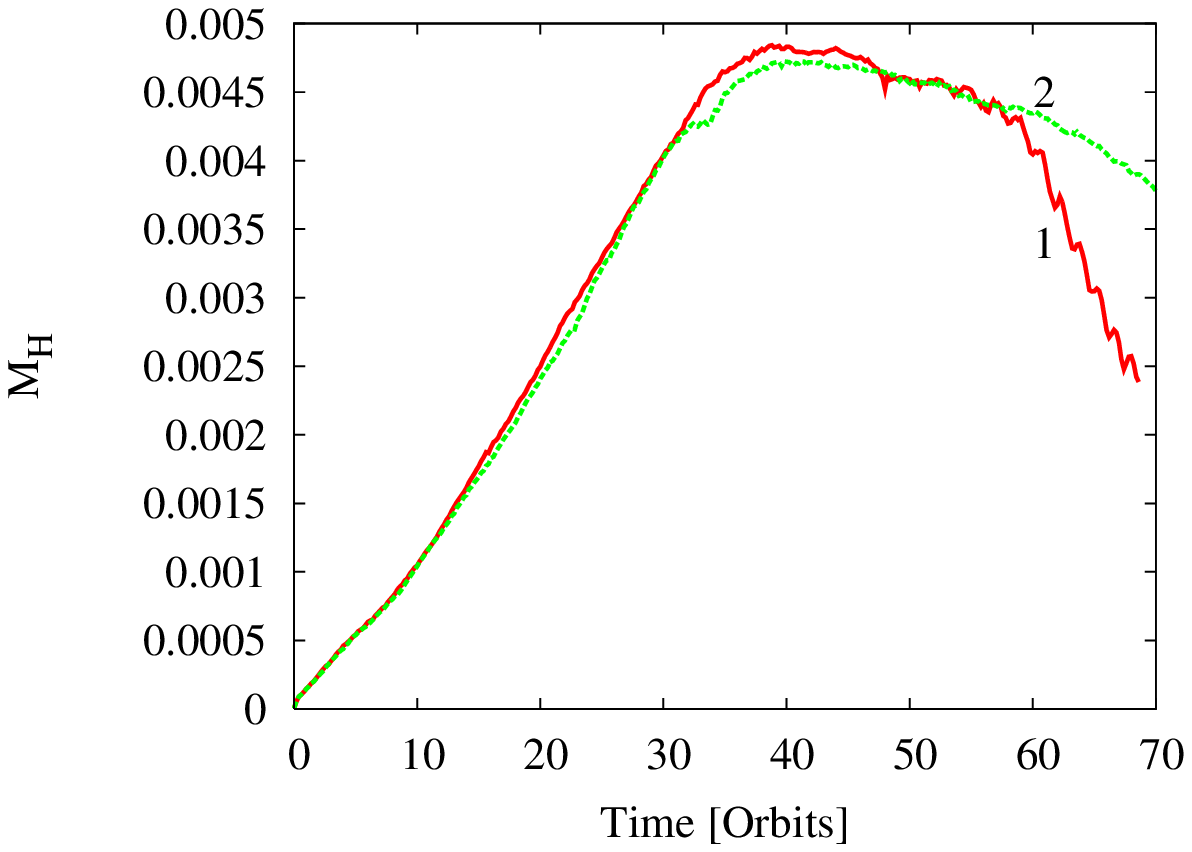}
\caption{Orbital evolution of the outward migrating planet for
different smoothing lengths of the planet's gravitational potential. 
The time evolution of the planet's semi-major axis $a$ (left panel)
and the mass of the gas inside a Hill sphere $M_H$ (right panel) are
plotted. Curves 1, 2
correspond to $r_\rmn{soft}$ equal $0.0208$ and $0.3 R_\rmn{H}$ for
$h_\rmn{p} = 0.4$. The planet mass $M_\rmn{P} = 0.001$.}
\label{f19_a_soft_4l}
\end{figure*}

\begin{figure*}
\includegraphics[width=84mm]{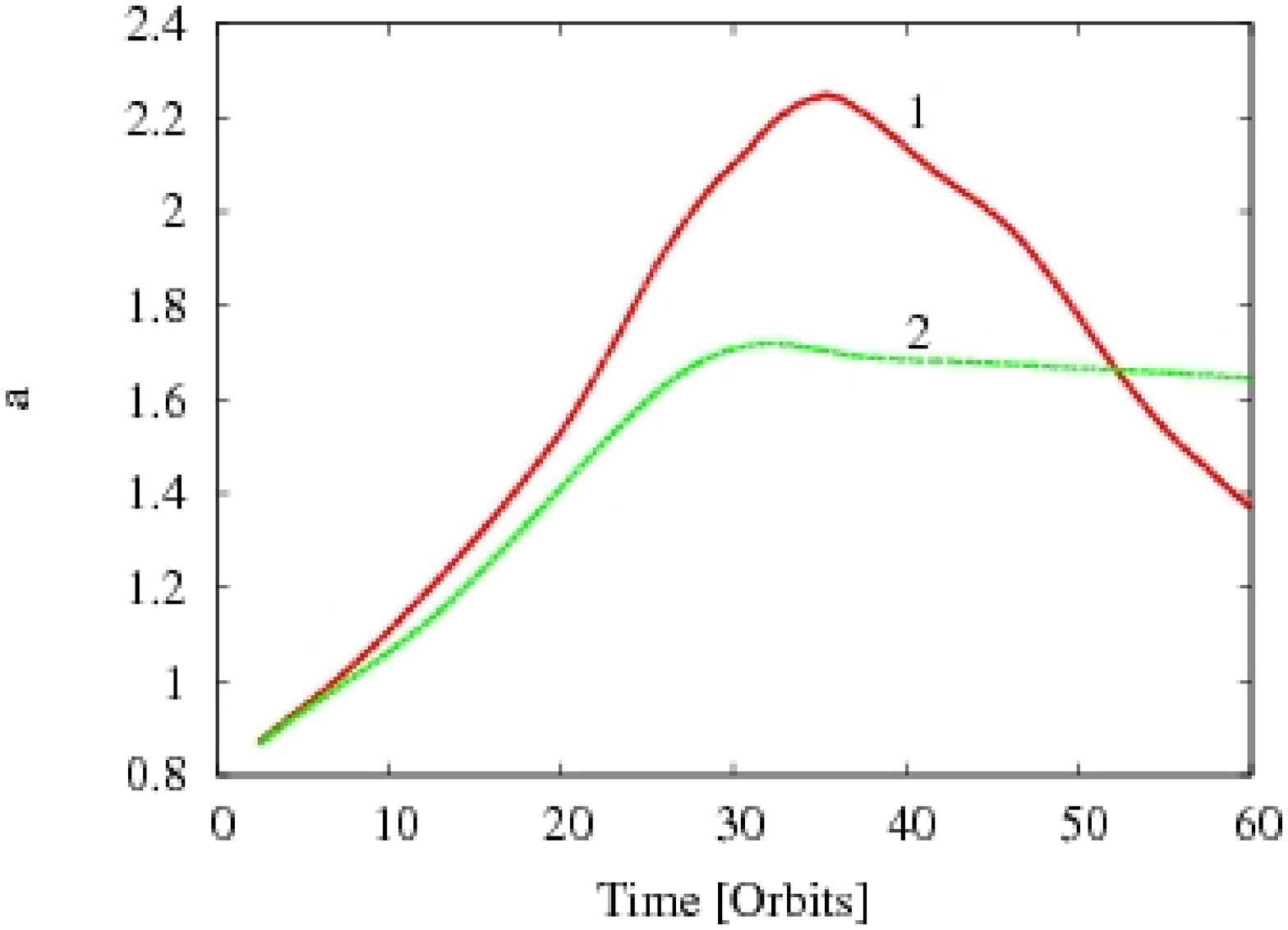}
\includegraphics[width=84mm]{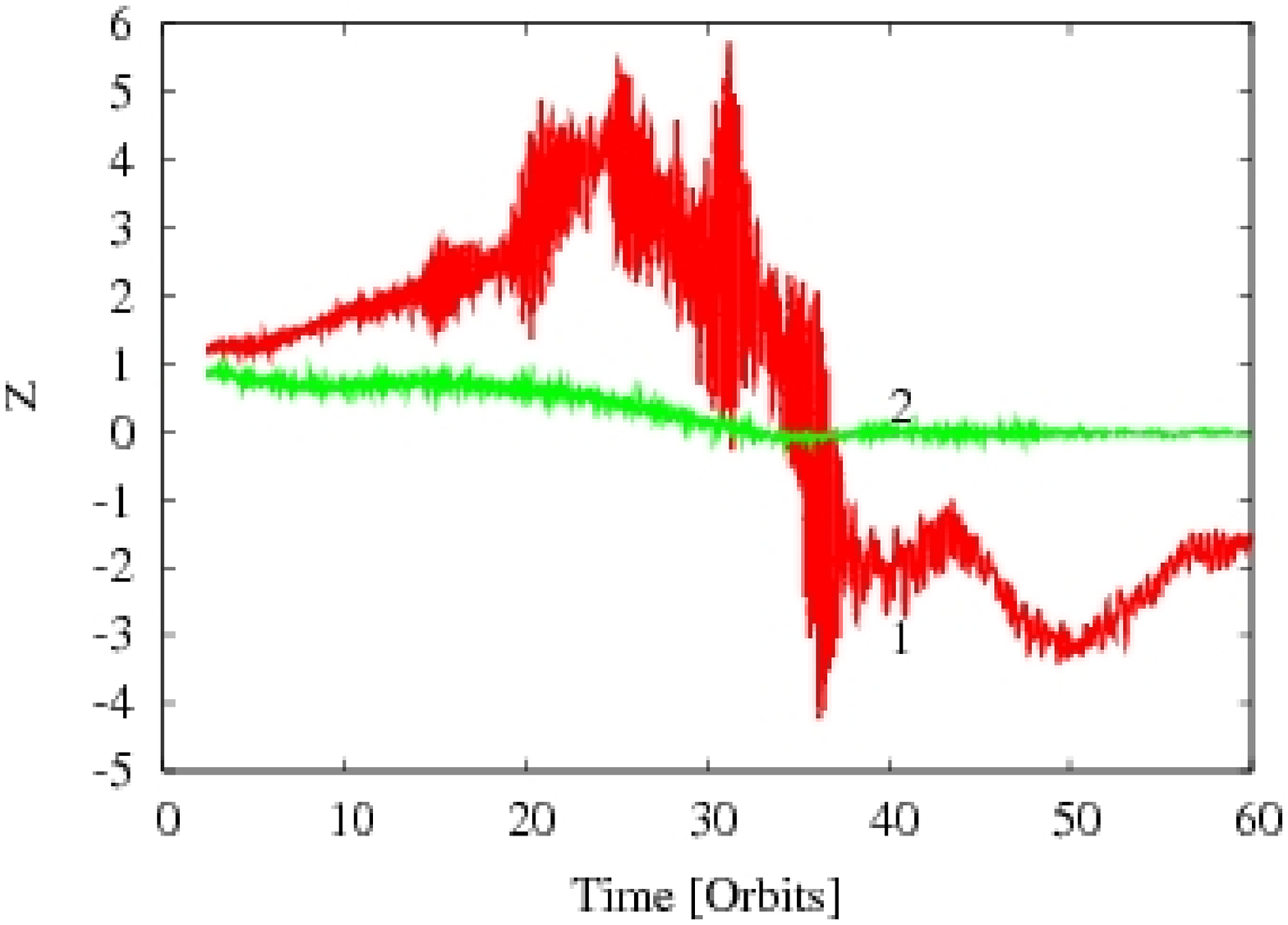}
\includegraphics[width=84mm]{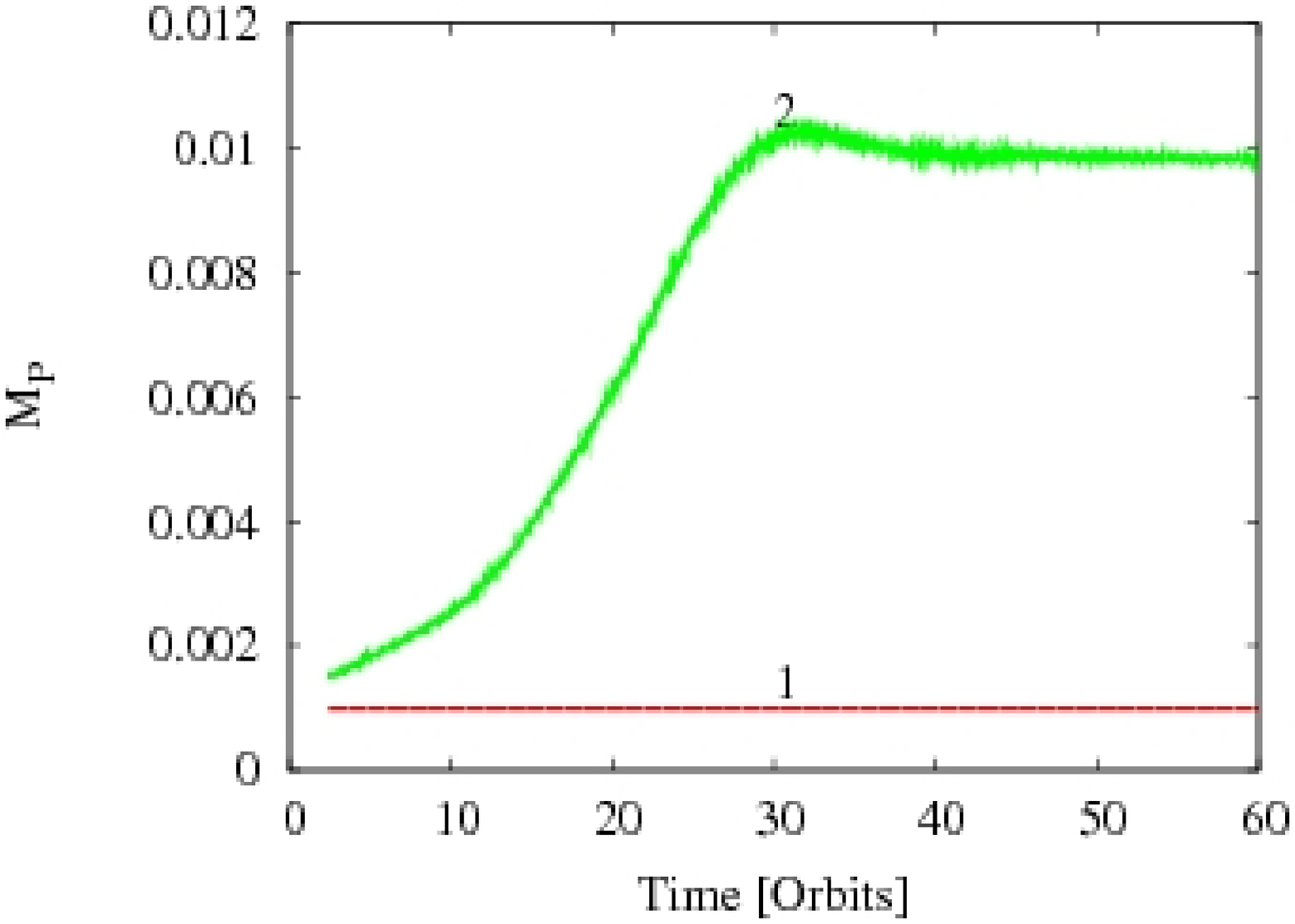}
\includegraphics[width=84mm]{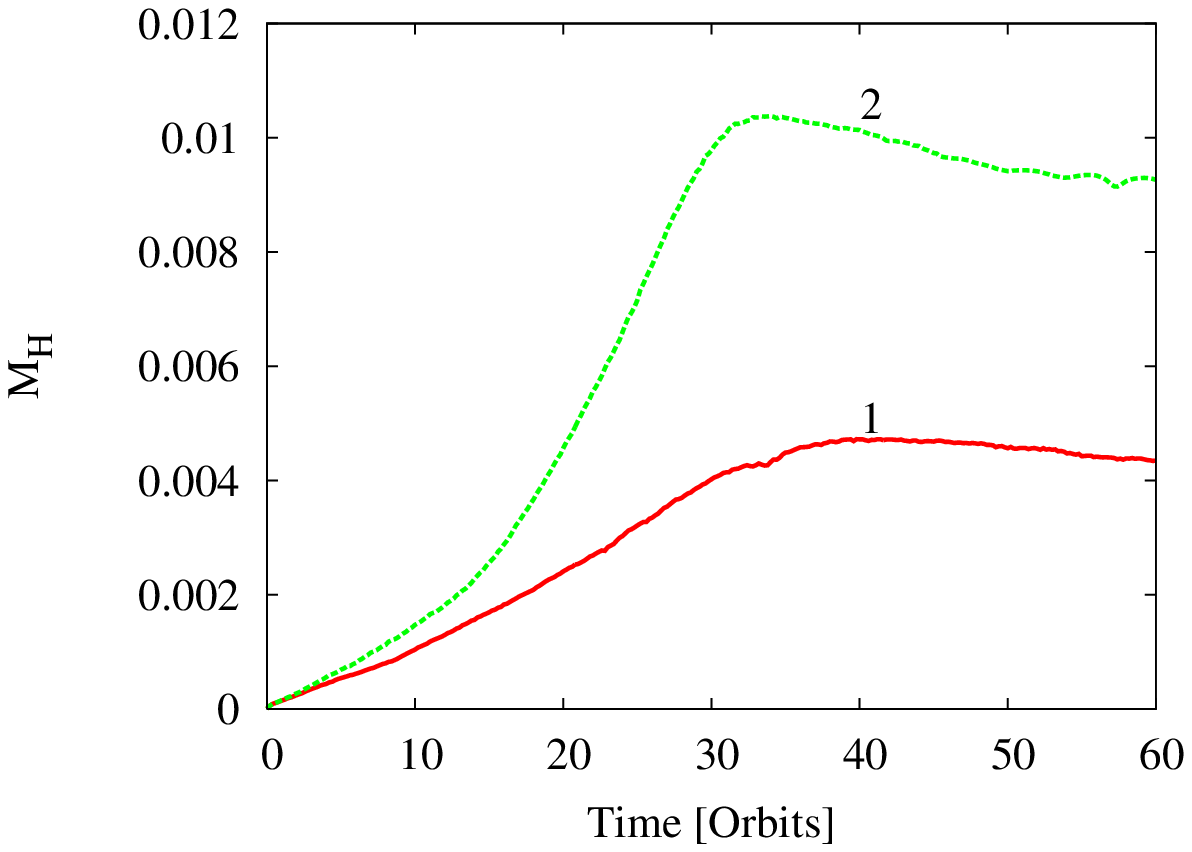}
\caption{Results of the simulations of outward migrating planet with
the varying planet's mass. Curve 1 corresponds to the model with the
constant planet's mass $M_\rmn{P} = 0.001$. Curve 2 shows the
simulation with the effective planet mass increased by the mass
within the planet's smoothing length. Upper left and upper right
panels show the orbital evolution and the non-dimensional migration
rate $Z$ for the inward migrating planets respectively. The planet
mass and the mass of the gas inside a Hill sphere are presented on the
lower left and lower right panels. The migration is sensitive to the
value of $\widetilde M_\rmn{P}$. In the first simulation the phases of
the outward and then inward directed rapid migration are visible, but
the second models shows only the first phase. Instead the planet is
locked in the relatively slow migration mode at around
$a=1.7$. However, during the first $20$ orbits both systems evolve in
the same way, even the the mass accumulated in the planet's proximity
differ by a factor of three.}
\label{fmin_d_planet_mass_out}
\end{figure*}

\subsection{Outward migration case}

\label{numerical_convergence_out}

Type III migration has no predefined direction and can result in both
inward and outward directed migration. In the previous section we
discussed the inward migration case. Most of the conclusions reached
there are also valid for the outward directed migration, even though
there are clear differences between the two types of simulations.

The first important difference is that the amount of mass accumulated
in the circumplanetary disc is much larger in the case of outward
migration, up to several times the initial planet mass. This is caused
by the combination of a different gas density at the planet's initial
position, and the fact that the Roche lobe size (and the relative grid
resolution) grows in the case of of outward migration.  In this
section we focus on the dependence of the planet's migration on the
value of the planet's gravitational softening and the treatment of
$M^*_\rmn{P}$.

\subsubsection{Gravitational softening}

\label{conv_grav_soft_out}

In Section~\ref{conv_grav_soft} we presented the dependence of the
planet's migration on $r_\rmn{soft}$ for the inward migration case. We
performed similar tests for the outward migrating planet.
Figure~\ref{f19_a_soft_4l} shows the planet's
orbital evolution and gas mass inside the Hill sphere for two
simulations with $r_\rmn{soft}$ equal $0.0208$ and $0.3 R_\rmn{H}$
respectively. In both simulations we used $h_\rmn{p} = 0.4$, and 
$M^*_\rmn{P}=M_P$.

In both cases the planet initially migrates outward but after about
$35$ orbits the direction of migration suddenly changes. We find that
the planet's orbital evolution is independent of $r_\rmn{soft}$ during the
first $25$ orbits. When the migration speed starts to slow down, some
differences between the two cases appear, and after the migration
direction has reversed a clear dependence on the smoothing length
becomes apparent. Still, the migration rate in both simulations
differs by less than than 20\% (when comparing $\dot a$ at a given
radius rather than at given time) and the mass of the gas inside the
Hill sphere remains similar up to $60$ orbits. The origin of the
differences is an instability of the horseshoe region that
arises for relatively high values of the migration 
rate. While the details of the instability 
are not entirely clear (streaming instability of gas with different vorticity
or alternatively accelerating/decelerating migration 
of a planet might be involved), the net result is a decrease of the 
mass deficit (density contrast) of the librating disk region, which 
decreases the corotational torque. During the migration reversal,
the numerically computed migration rate depends strongly on $r_\rmn{soft}$.  
Therefore we limit our investigation in the case of
$M^*_\rmn{P}= M_\rmn{P}$ to the outward migration phase only, which is
independent of $r_\rmn{soft}$.

The dependence of the torques $\Gamma_\rmn{soft}$, $\Gamma_\rmn{out}$
and $\Gamma_\rmn{RL}$ on $r_\rmn{soft}$ is similar for both the inward
and the outward migration.

\subsubsection{Mass accumulation in the planet's vicinity}
\label{conv_grav_mass_out}

In Section~\ref{conv_grav_mass_in} we presented models for an inward
migrating planet where we used different prescriptions for
$M^*_\rmn{P}$. In the case of the outward migrating planet we
performed similar simulations with $r_\rmn{soft} = 0.3 R_\rmn{H}$,
$h_\rmn{p} = 0.4$ and $\mu_\rmn{D}=0.01$. Here we only consider the
cases of $M^*_\rmn{P}=M_\rmn{P} = 0.001$ and $M^*_\rmn{P}=\widetilde
M_\rmn{P}$, i.e.~we do not consider the accretion case.  We present
the results in Fig.~\ref{fmin_d_planet_mass_out}.  Curves 1 and 2
correspond to the $M_\rmn{P}$ and the $\widetilde M_\rmn{P}$ case respectively. The
upper left and upper right panels show the orbital evolution and the
non-dimensional migration rate $Z$, the planet mass and the mass of
the gas inside the Hill sphere are presented on the lower left and lower
right panels.

The orbital evolution of the outward migrating planet is sensitive to
the choice for $M^*_\rmn{P}$. In the $M^*_\rmn{P}=M_\rmn{P}$
simulation the planet reaches $a = 2.25$ and rapidly changes its
direction of migration (as above). For the $M^*_\rmn{P}=\widetilde
M_\rmn{P}$ case the mass accumulated in the planet's proximity grows
from an initial value of about $4.5$ ($M_\rmn{P}=0.001$) up to about
$10.5$ Jupiter masses. This behaviour is similar to the case of inward
migrating planet with $h_\rmn{p} = 0.3$.  The increase of
$M_\rmn{soft}$ causes the planet to stop around $a=1.75$ after $32$
orbits, since the disc is not massive enough to support rapid
migration of such a high mass object. 

The effects of neglecting the gravitational interaction between the
massive planet's envelope and the whole disc are visible on the plot
of the non-dimensional migration rate $Z$. In the
$M^*_\rmn{P}=M_\rmn{P}$ simulation the planet migrates in a fast
migration limit with $Z > 1$ and reaches $Z=4$ before it changes its
direction of migration. For $M^*_\rmn{P}=\widetilde M_\rmn{P}$ $Z$
stays below $1$ and the planet migrates in the slow migration limit.
However, the big difference in $Z$ is caused mostly by the different
planet's effective mass. During first $20$ orbits the semi-major
axis $a$ differs by less than $7\%$ and the migration rate $\dot a$
differs at most by $25 \%$. During this initial period both systems
evolve in the same way, even though the mass accumulated in the
planet's proximity differs by a factor of three (about $2$ and $6$
$M_{\jupiter}$ for the two models after $20$ orbits). This is caused
by the fact that for $Z>1$ the migration rate depends only weakly
on the planet's mass, and during the first $20$ orbits $Z$ is close
to $Z=1$ in the $M^*_\rmn{P}=\widetilde M_\rmn{P}$ simulation.

This test shows that it is important to use $\widetilde M_\rmn{P}$ 
when we want to study the slow migration limit of migration type III 
and its stopping. However, just using a constant $M_\rmn{P}$ gives
a correct description of the fast migration limit. 
For $h_\rmn{p} = 0.6$ the amount of mass accumulated in the circumplanetary 
disc and the planet's migration become independent on the treatment 
of the effective mass.
A more detailed description of the outward-directed type III migration
will be given in Paper~III.


\section{Conclusions}

\label{conclusions}

We investigated the orbital evolution of a Jupiter-mass planet
embedded in a disc and the dependence of its migration on the adopted
disc model, as well as the representation of the disk-planet coupling. 
The calculations were performed in two dimensions in an
inertial frame of reference, using a Cartesian coordinate system.  We
focused especially on the details of the flow in the planet's
proximity. We employed an adaptive mesh refinement code with up to 5
levels of refinement which provides high resolution in the planet's vicinity. We conducted 
a careful analysis of the effects of various numerical
parameters that influence the flow patterns inside the Roche lobe and
the dependence on the grid resolution. Our goal was to find a disc
model that allows numerical convergence to be reached and that
minimises non-physical effects such as an artificial increase of the
planet's inertia. We introduced a first-order correction of the gas
acceleration in the circumplanetary disc due to the gas self-gravity
and a modification of the local isothermal approximation, which
allows us to take into account the gravitational field of the planet
when calculating the local sound speed.

For a simple disc model without these modifications to the temperature
profile and the gas acceleration, we obtained results consistent with
those of \citet{2005MNRAS.358..316D}, depending strongly on resolution and gravitational smoothing. Convergence could
only be achieved if the torques from the interior of the Hill
sphere $\Gamma_\rmn{RL}$ are neglected. However, the interior of
the Roche lobe is important for driving rapid migration and cannot be
neglected when numerically studying type III migration.  
Since this region is complex and dynamically variable, 
it is impossible to define a unique radius around the planet 
that would separate the planet's sphere of influence 
from the disk, and thus could be neglected in the torque calculation.
Therefore, one has to include all the torques from the Roche
lobe, even though artificial gravitational softening must be employed on 
spatial scale which is much smaller than the Roche lobe. 
Moreover, in the case of fast migration, the interior of the
Roche lobe is not separated from the corotational flow and we cannot
neglect the gas inflow into the circumplanetary disc.

When including the torques from the entire Roche lobe, we achieve
convergence when using the modification of the local isothermal
approximation and the correction for self-gravity. Both corrections
are needed and we need to use a scale height of the circumplanetary
disc $h_\rmn{p} \ge 0.4$ to achieve this convergence. In this case the
results are insensitive to the gravitational smoothing length
$r_\rmn{soft}$.

However, the (converged) solution will depend on the choice for
$h_\rmn{p}$.
It determines the flow structure inside the Roche lobe
and the amount of gas accumulated in the planet's vicinity. The latter
influences the migration behaviour as the accumulation of a large mass
lowers the non-dimensional migration rate $Z$. As long as
$|Z|>1$, the planet migrates rapidly at a rate that is only weakly
dependent on the planet's mass and the conditions in the
circumplanetary disc. If the mass accumulation forces $|Z|$ to drop below
1, migration will slow down and the planet may  
be locked in the disc in a mode resembling type II migration where, strictly
speaking, corotational torques may still be dominant over Lindblad torques.

Our modifications of the disc model compared with previous investigations
are physically justified,
since the rapid accretion during migration will release potential energy and 
heat the gas in the
planet's environment, while self-gravity will start to play a role if
the planet collects an amount of mass of order its original mass.
However, our implementation of these two effects are admittedly still very 
crude approximations. Ideally, non-isothermal, self-gravitating models
should be used to self-consistently study type III migration. For
example, the assumption that the circumplanetary disk aspect ratio 
$h_\rmn{p}$ is constant during the whole
simulation is probably not realistic. However, type III migration is
only weakly dependent on the mass accumulation in the fast migration limit
($|Z|>1$), so we expect variations of $h_\rmn{p}$ to play an
important role only in the slow migration limit ($|Z|<1$). 

In summary, the migration rates of the freely migrating planet in a
massive disc (satisfying the condition for rapid migration) are
sensitive to the adopted disc model, especially to the treatment of
the interior of the Roche lobe. The effects of heating close to the
planet and self-gravity of the gas have to be taken into account to construct consistent
models for type III migration. These models then show rapid migration
independent of the numerical resolution used. In Papers~II and~III we will 
explore the physics of type III migration in more detail.


\section*{Acknowledgements}

The software used in this work was in part developed by the
DOE-supported ASC / Alliance Center for Astrophysical Thermonuclear
Flashes at the University of Chicago. We thank F.Masset for
interesting and useful comments. Some of the calculations used the
resources of High Performance Computing Centre North (HPC2N) and
National Supercomputing Centre, Link\"oping.  AP acknowledges
financial support provided through the European Community's Human
Potential Programme under contract HPRN-CT-2002-00308, PLANETS.

%
%
\bibliographystyle{mn2e}
\bibliography{articles_astro}
%
\appendix

\section[]{Numerical diffusivity of code}
\label{app_num_diff}

Our code participated
in the {\it Hydrocode Comparison for a Disc-Planet System Project}
\citep{2006MNRAS.370..529D}. In this project the results of a wide
range of hydrodynamics codes were compared for the case of Jupiter-
and Neptune- mass planets on a constant orbit. In this comparison
differences between our Cartesian and the other (mostly) cylindrical
codes were found, including a version of {\it FLASH} using a
cylindrical grid.

The main difference between the Cartesian and cylindrical {\it FLASH}
(denoted (FLASH-AP and FLASH-AG respectively in
\citep{2006MNRAS.370..529D}) is a depletion of the inner disc. This
leads to a deeper and broader gap, an asymmetry between the tadpole
regions around the L4 and L5 Lagrangian points and a difference in the
differential Lindblad torque calculation. The depletion is an effect
of mass accumulation in the star's vicinity, which takes place in a
poorly resolved part of the disc (compared to the cylindrical codes).
Higher resolution simulations (not discussed in
\citealt{2006MNRAS.370..529D}) show that this behaviour depends strongly
on the grid resolution.

\begin{figure}
\includegraphics[width=84mm]{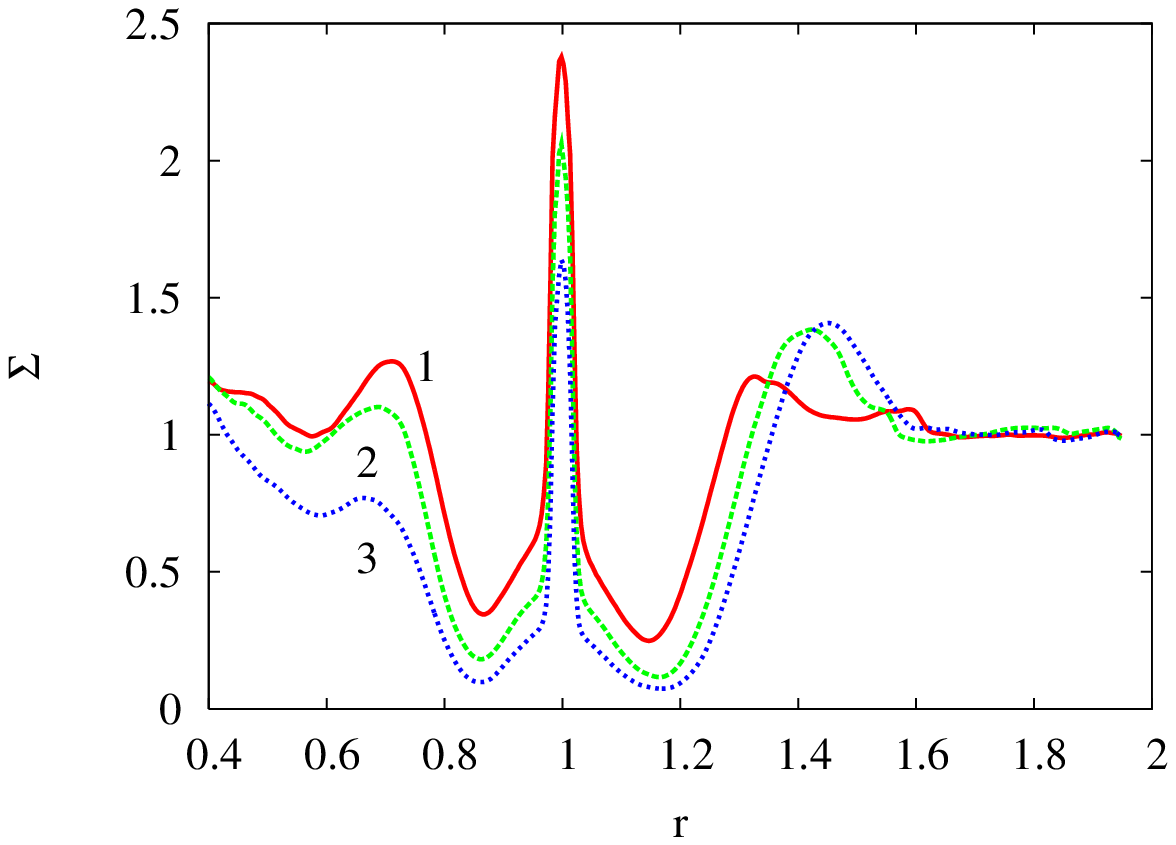}
\caption{The normalised surface density profiles averaged azimuthally
over $2 \pi$ for the Jupiter-mass planet. The star's neighbourhood is
not plotted to make the plot similar to Fig.~4 in
\citet{2006MNRAS.370..529D}. The curves 1, 2 and 3 correspond to 60,
150 and 300 orbits.}
\label{fig_comp_t_jup}
\end{figure}

To further investigate this feature and its relevance for the
simulations presented in the paper, we performed a test employing the
typical mesh size adopted in our investigation. We used
$\alpha_\Sigma=0$ and the Jupiter-mass planet is kept on a constant
orbit with semi-major axis $a=1$. The disc extends from $-2.0$ to
$2.0$ in both directions around the centre of mass and the grid has
$800$ cells in both directions. The normalised surface density profile
averaged azimuthally over $2 \pi$ for $60$, $150$ and $300$ orbits are
presented in Fig.~\ref{fig_comp_t_jup}
\footnote{ The results for {\it Hydrocode Comparison for a Disc-Planet
System Project} for different times and resolution are presented at
http://www.astro.su.se/groups/planets/comparison/codes.html}. As
type III migration for Jupiter in our simulations takes
about $60$ orbits on average, curve 1 is the most
relevant. At the higher resolution the normalised surface density
profile is almost identical to the results of the other codes
in \citet{2006MNRAS.370..529D}.  The depletion of the inner disk only
appears at later times (curves 2 and 3).  This means that the gas
accumulation in the star's vicinity is a slow process and does not
play an important role for type III migration of a Jupiter-mass
planet, which is anyway dominated by the co-orbital torques. Clearly
for a study of the slower Type II migration, dominated by the Lindblad
torques, this effect is less than beneficial.

\label{lastpage}

\end{document}